\documentclass{aa}  

\usepackage{graphicx}
\usepackage{array}
\usepackage{txfonts}
\usepackage{caption}
\usepackage{lipsum}
\usepackage{hyperref}
\usepackage{url}
\usepackage{natbib}
\usepackage{breqn}
\usepackage{diagbox}
\usepackage{cuted} 
\usepackage[mathscr]{euscript}

\newcommand{\iu}{{i\mkern1mu}}

\newcolumntype{C}{>{\centering\arraybackslash}p{0.2\textwidth}}
\usepackage{enumitem}
\usepackage{booktabs}
\bibpunct{(}{)}{;}{a}{}{,} 
\begin{document} 
\title{Inferring magnetic field strengths in high-redshift lensing galaxies from Faraday rotation measure observations}
\author{Stefan Reissl \inst{\ref{inst1}}, Ralf S. Klessen \inst{\ref{inst1},\ref{inst2}}, Annalisa Pillepich \inst{\ref{inst5}} , Sui Ann Mao \inst{\ref{inst6}},   R{\"u}diger Pakmor \inst{\ref{inst7}}, \and Dylan Nelson \inst{\ref{inst1}} }

\authorrunning{Reissl et al.} 

\institute{
\centering Universit\"{a}t Heidelberg, Zentrum f\"{u}r Astronomie, Institut f\"{u}r Theoretische Astrophysik, Albert-Ueberle-Straße 2, \\  D-69120 Heidelberg, Germany \label{inst1} 
\and
\centering Universit\"{a}t Heidelberg, Interdisziplin\"{a}res Zentrum f\"{u}r Wissenschaftliches Rechnen, Im Neuenheimer Feld 205, \\ D-69120 Heidelberg, Germany  \label{inst2}
\and
\centering Max-Planck-Institut f\"{u}r Astronomie, Konigstuhl 17, D-69117 Heidelberg, Germany  \label{inst5}
\and
\centering Max-Planck-Institut f\"{u}r Radioastronomie, Auf dem H\"{u}gel 69, D-53121 Bonn, Germany  \label{inst6}
\and
\centering Max-Planck-Institut f\"{u}r Astrophysik, Karl-Schwarzschild-Str. 1, D-85748, Garching, Germany  \label{inst7}
}
			
\date{}

\abstract{Observations of Faraday rotation in the lensing-galaxy of \textsc{CLASS B1152+199} at $z=0.439$ suggest that efficient dynamo mechanisms had already amplified galactic magnetic fields by this epoch. In this work, we evaluate the reliability of this method for deriving magnetic field strengths and assess the likelihood of observing Faraday rotation differences similar to those measured in \textsc{CLASS B1152+199}. We select 64 star-forming disk galaxies from the TNG50 cosmological magnetohydrodynamical simulation analog to \textsc{CLASS B1152+199} in mass and redshift and forward model these systems using polarized radiative transfer calculations  with POLARIS and subsequent RM synthesis.We characterize the large-scale magnetic fields directly from the TNG50 galaxies by spatially averaging the magnetic
field components and fitting their radial profiles. TNG50 predicts characteristic field strengths of
$B_0=1.9-15.6\ \mu\mathrm{G}$ for the individual galaxies,
while fitting the combined sample yields
$B_0=10.3\ \mu\mathrm{G}$. The magnetic field strength $B_{\mathrm{obs}}$ inferred from the synthetic observations generally recovers $B_{\mathrm{0}}$ within a factor of three. We find only a minor dependence of $B_{\mathrm{obs}}$ on the assumed magnetic field geometry, while galaxy inclination and beam depolarization have a stronger impact on the inferred field strength. The differential rotation measure recovered from RM synthesis remains correlated with the directly integrated value, although beam depolarization systematically reduces its magnitude. Finally, only $1.41\%$ of the directly integrated sightline pairs reproduce the observed $\Delta\mathrm{RM}=1040\pm60\ \mathrm{rad\,m^{-2}}$ of \textsc{CLASS B1152+199} assuming that contributions unrelated to the lensing galaxy cancel between the two images. This fraction increases to $3.16\%$ when restricting the sightlines to galactocentric distances comparable to the observed $2.6$ and $6.5\ \mathrm{kpc}$. We conclude that differential Faraday rotation of gravitationally lensed sources provides a robust probe of large-scale magnetic fields, while the large $\Delta\mathrm{RM}$ observed in \textsc{CLASS B1152+199} appears to be relatively rare among the considered TNG50 analog galaxies.}


\maketitle


\section{Introduction}

Magnetic fields are fundamental components of the Universe, influencing a wide range of astrophysical processes and systems \citep{Beck2015}. The presence of magnetic fields in the interstellar medium (ISM) has significant implications for galaxy evolution and star formation. They regulate the dynamics of the ISM by mediating turbulence, shaping molecular cloud structures, and providing pressure support against gravitational collapse \cite[e.g.][]{MacLow2004, McKee2007, Klessen2016}. These fields influence star formation efficiency by suppressing or promoting the collapse of gas clouds, depending on the local magnetic field geometry and strength \cite[e.g.][]{Federrath2012, Beck2013, Haverkorn2015, Federrath2015}. In addition, magnetic fields are essential for the acceleration and propagation of cosmic rays (CRs), which themselves affect ISM ionization and energy balance \citep{Zweibel1997}. Magnetic fields also interact with disk dynamics, affecting angular momentum transport and gas accretion onto galactic centers, which can influence the growth of supermassive black holes and nuclear starbursts.

Despite their importance, the origin of magnetic fields remains a challenging question. In the early Universe, weak seed fields may have emerged through mechanisms such as battery effects, turbulence during cosmic reionization, or magnetohydrodynamic instabilities in primordial environments \citep{Durrer2013}. These fields underwent amplification via small-scale turbulent dynamo processes, which efficiently increase magnetic energy in the chaotic environments of early galaxies \citep[][]{Schekochihin2004, Brandenburg2005, Federrath2011, Schober2012a, Federrath2014, Schober2015}. Over time, as galaxies evolved and developed ordered rotation, large-scale dynamo processes became dominant, generating the coherent magnetic fields observed in mature galaxies such as the Milky Way \citep{1996ARAA}.

Magnetic fields at high redshifts are also important for interpreting extragalactic observations. For instance, magnetic fields impact the interpretation of 21-cm emission during the epoch of reionization, complicating the study of early structure formation \citep{Furlanetto2006}. These effects underline the necessity of understanding the role of magnetic fields in the context of both local and cosmological environments.

Observations of the Milky Way and nearby galaxies provide
important constraints on the strength and structure of galactic
magnetic fields. In the solar neighborhood, the total magnetic
field strength is approximately $6\ \mu\mathrm{G}$, while the
regular component inferred from Faraday rotation is weaker,
with a strength of a few $\mu\mathrm{G}$ \citep{Beck2009,Beck2015}. For nearby star-forming galaxies, radio synchrotron observations typically indicate total magnetic field strengths of several to tens of $\mu\mathrm{G}$ \citep{Beck2015}. More recently, \citet{Heesen2023} used LOFAR observations at $144\ \mathrm{MHz}$ to estimate magnetic field strengths in 39 nearby galaxies, finding mean values between
$3.6$ and $12.5\ \mu\mathrm{G}$, with a sample average
of $7.9\pm2.0\ \mu\mathrm{G}$. They also found a strong correlation between magnetic field strength and total gas surface density, illustrating the connection between galactic magnetic fields and the interstellar medium. However, synchrotron-based field strengths generally rely on assumptions about cosmic-ray electrons and primarily constrain the total magnetic field, whereas Faraday rotation provides information on the coherent LOS component. Extending such measurements to higher redshifts is therefore important for constraining the evolution of large-scale galactic magnetic fields.

Observations of a galaxy at various redshifts $z$ have provided valuable insights into the evolution of magnetic fields \citep[e.g.][]{Geach2023}. In the pioneering observation of \cite{Mao2017NatAs}, the detection of a micro-Gauss large-scale magnetic field in a galaxy is reported for now the most distant galaxy to date with $z=0.439$. Here, the field properties within a lensing galaxy are probed using the difference of Faraday rotation of about $\Delta \mathrm{RM} = 1040\ \mathrm{rad\ m^{-2}}$ between two distinct lensed images at impact parameters of $2.6\ \mathrm{kpc}$ and $6.5\ \mathrm{kpc}$. The lensed images originate from a polarized background source positioned at a redshift of $z=1.019$ \citep{Myers1999,Rusin2002}. The recovered large-scale magnetic field strength of the lensing galaxy is estimated to be in the range ${ 4\ \mu\mathrm{G} - 16\ \mu\mathrm{G} }$ for an assumed axisymmetric morphology and ${ 7\ \mu\mathrm{G} - 17\ \mu\mathrm{G} }$ for a BS one \citep[see e.g.][]{Beck2007,Mao2017NatAs}. Consequently, these synchrotron polarization and Faraday rotation measures indicate that star-forming galaxies at redshifts of about $z = 0.4 - 0.5$ already host magnetic fields with strengths comparable to those of present-day galaxies \citep{Beck2015}.

These observations challenge views of magnetic field amplification as a slow process and suggest that efficient dynamo mechanisms were already at work in the early Universe \citep{Ruzmaikin1988, Kulsrud2008, Schleicher2010, Sur2010, Schober2012b, Schober2013}. Hence, the understanding of magnetic fields at high redshifts is essential for addressing fundamental questions about galaxy formation and evolution. While the majority do not, some recent cosmological simulations incorporate magnetohydrodynamics (MHD) to model magnetic field amplification and its impact on galaxies across cosmic time \citep{Pakmor2014,Pakmor2024}. These simulations, coupled with advanced observational techniques like polarized radiative transfer and high-resolution radio interferometry, are paving the way for a more comprehensive understanding of the cosmic magnetic field's evolution and its broader implications for the Universe's structure and dynamics.

In this paper, we evaluate the reliability of the method utilized in \cite{Mao2017NatAs} to derive the large-scale magnetic field strength in a lensing galaxy from the polarized radiation of a background source. To this end, we use a selected sample of TNG50 galaxies as analogs to model the lensing galaxy in the \textsc{CLASS B1152+199} system.
Furthermore, we explore how likely it is to find observational
constraints on the magnetic field strength comparable to those obtained for \textsc{CLASS B1152+199} \citep{Myers1999,Rusin2002}.

This paper is structured as follows: In Sect.\ref{sect:TNG50}, we introduce the TNG50 cosmological simulation and the selection of galaxies similar to the lensing galaxy in \textsc{CLASS B1152+199}. Our sample selection analysis methodology is outlined in Sect.\ref{sect:TNG50samples}. In Sect.\ref{sect:synthobs} we review the basic physical processes we utilize to mimic synthetic synchrotron emission and Faraday rotation by means of RT calculations. The selection of properties for the background sources is introduced in Sect.\ref{sect:source}. In Sect.\ref{sect:lensing}, Sect.\ref{sect:FaradaySynthesis}, and Sect.\ref{sect:mag_est}, we outline the calculation of lensing points, the Faraday rotation measure (RM) synthesis method, and the magnetic field strength estimation from synthetic observations, respectively. The resulting dependence on inclination, beam depolarization, and a synchrotron-active lensing galaxy are presented in Sect.\ref{sec:inclination}, Sect.\ref{sec:depol}, and Sect.\ref{sec:sync}, respectively. In Sect.\ref{sec:distribution}, we evaluate the probability of finding the exact Faraday rotation observed in \textsc{CLASS B1152+199}. Finally, in Sect.\ref{sec:summary}, we summarize our findings.


\section{Methods}

\subsection{The TNG50 Simulation} \label{sect:TNG50}

We analyze galaxies from the TNG50 cosmological simulation \citep{Nelson2019,Pillepich2019}, the highest resolution volume of the IllustrisTNG simulation suite \citep{Springel2018,Pillepich2018,Nelson2018,Naiman2018,Marinacci2018}. The TNG50 simulation combines high spatial and mass resolution with a representative cosmological volume, TNG50 evolves a periodic cube with a sidelength of ${51.7\ \mathrm{cMpc} }$. It uses the moving-mesh code \textsc{Arepo}, which solves the coupled equations of self-gravity with ideal magnetohydrodynamics (MHD) an adaptive Voronoi grid \citep{Springel2010}. 

TNG50 traces the evolution of $2 \times 2160^3$ dark matter particles and gas cells from redshift ${ z=127 }$ to ${ z=0 }$. Its numerical resolution gives a baryonic mass of ${ 8.5 \times 10^4\ M_\odot }$ and ${ 4.5 \times 10^5\ M_\odot }$ for dark matter. Spatially this gives gas cell resolutions of of ${ 70\ \mathrm{pc} - 140\ \mathrm{pc} }$ within the star-forming regions of each galaxy. The simulation incorporates the fiducial TNG galaxy formation model \citep{Weinberger2017,Pillepich2018a}. This includes star formation, stellar evolution, chemical enrichment, supernova (SN) feedback, and active galactic nucleus (AGN) feedback, as well as magnetic field evolution. 

The magnetic field is evolved self-consistently starting from a vanishingly small primordial seed field. It begins with uniform amplitude in a single direction, and then evolves due to advection, amplification, and dissipation \citep{Pakmor2017}. The resulting field strengths and morphologies at low redshifts are broadly consistent with observations of local galaxies, reproducing the amplification of fields through dynamo processes in star-forming disks. The field within galaxies is typically of the order of $1\ \mu\mathrm{G}$ to $10\ \mu\mathrm{G}$, depending on galaxy mass and type \citep{Nelson2018,Pillepich2024}. It is weaker at lower density \citep{Marinacci2018}, declining into the  circumgalactic medium \citep[CGM;][]{Ramesh2023} and producing large size albeit weak field strength bubbles in the intergalactic medium \citep{Aramburo2021}. While the inclusion of magnetic fields has an overall minor impact on galaxy formation in the TNG model \citep{Pillepich2018a}, it can reach equipartition within the interstellar medium and even dominate over thermal pressure support in cool CGM gas, for example \citep{Nelson2020,Ramesh2024}.

In TNG50 halos are identified using the friends-of-friends algorithm, and then subhalos and galaxies are identified with the \textsc{subfind} algorithm \citep{Springel2001}.

Finally, to simplify our later post-processing and analysis, each selected TNG50 galaxy is rotated into a face-on orientation. For this purpose, we determine the minor axis from the shape tensor of the stellar particle mass distribution within the stellar half-mass radius. All relevant vectors, including coordinates, velocities, and magnetic field vectors, are then rotated such that this minor axis is aligned with the $z$-axis. We emphasize that the orientation is therefore determined from the stellar mass distribution rather than from the angular momentum of the stellar or gaseous component. This procedure ensures a well-defined orientation, with the LOS
parallel to the aligned stellar minor axis defining the face-on direction, corresponding to an inclination of $i=0^\circ$.

\begin{figure*}[]
	\centering
	\includegraphics[width=1.0\textwidth]{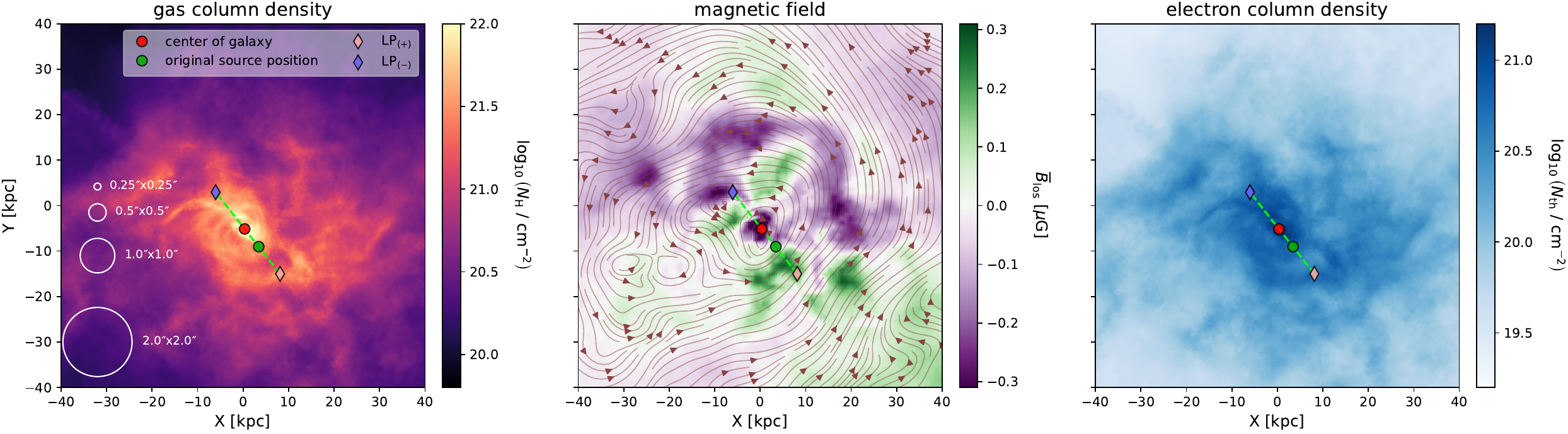}
	\caption{Left panel: Face-on map of the gas column density $N_{\mathrm{H}}$ of the exemplary lensing galaxy ID-1026 with a redshift of  $z=0.4$ from our TNG50 simulation sample. The red dot marks the peak value of  $N_{\mathrm{H}}$, and the green dot indicates the true offset position of a fictitious lensed galaxy behind. Purple and pink diamonds represent the corresponding positive (LP$_{(+)}$) and negative (LP$_{(-)}$) solutions of the lensing equation for the background galaxy. White circles indicate the considered beam sizes in arcseconds. Midle panel: The same as the left panel, but for the magnetic field in the disk averaged along the line of sight (LOS). Color coded are magnitude and sign of the LOS component $B_{\mathrm{LOS}}$. Streamlines depict the field component in the disk perpendicular to $B_{\mathrm{LOS}}$. Right panel: The column density map $N_{\mathrm{th}}$ of the distribution of thermal electrons, corrected by star formation as described in the text. }
	\label{fig:InputParams}
\end{figure*}

\subsection{Sample Selection of TNG50 galaxies}
\label{sect:TNG50samples}
In order to mimic the lensing system \textsc{CLASS B1152+199}, we identify 64 star-forming galaxies with a star formation rate (SFR) of
${10^{-2}\ M_\odot\ \mathrm{yr}^{-1} - 2\ M_\odot\ \mathrm{yr}^{-1}}$
and a half-mass radius of ${r_{\mathrm{1/2}}=20\ \mathrm{kpc} - 40\ \mathrm{kpc}}$ for the gas component. The selected galaxies have a total mass of ${7\times10^{10}\ M_\odot - 2\times10^{12}\ M_\odot}$,
corresponding to a gas mass of ${9\times10^{9}\ M_\odot - 2\times10^{10}\ M_\odot}$ and a stellar mass of
${2\times10^{8}\ M_\odot - 3\times10^{9}\ M_\odot}$. This selection covers a broad range of galaxy masses and SFRs to account for the uncertain properties of the lensing galaxy in \textsc{CLASS B1152+199}
\citep{Myers1999,Toft2000,Rusin2002,Dai2009}. In particular, the mass and gas distribution of the lensing galaxy are not sufficiently constrained to establish a close correspondence with individual TNG50 galaxies. If the actual lensing galaxy is substantially more massive than most galaxies in our sample, our selection may systematically underestimate the differential rotation measures expected for comparable lines of sight. Consequently, the selected galaxies should be regarded as a sample of plausible analogs rather than exact counterparts of
the observed lensing galaxy. We select galaxies from $z=0.4$
as well as $z=0.5$. For simplicity, we apply a distance corresponding to $z=0.5$ for the lensing galaxy in the following sections. Column density maps for all 64 selected TNG50 galaxies are provided in Appendix~\ref{app:maps}.

\subsection{Extracting magnetic field properties from TNG50 galaxies}
For non-star-forming cells we utilize the number density of thermal electrons $n_{\mathrm{th}}$ as a fraction of the hydrogen number density. However, star-forming gas resides on the effective two-phase ISM equation of state, such that the majority of the mass is cold and does not contribute to RM because of its low volume-filling fraction and therefore
small probability of intersecting the LOS \citep{Pakmor2018}. Hence, we use only the fraction of the total gas mass that is in the hot volume filling and fully ionized phase for our RT calculations \citep{Springel2003}.

\begin{figure*}[]
	\centering
	\includegraphics[width=1.0\textwidth]{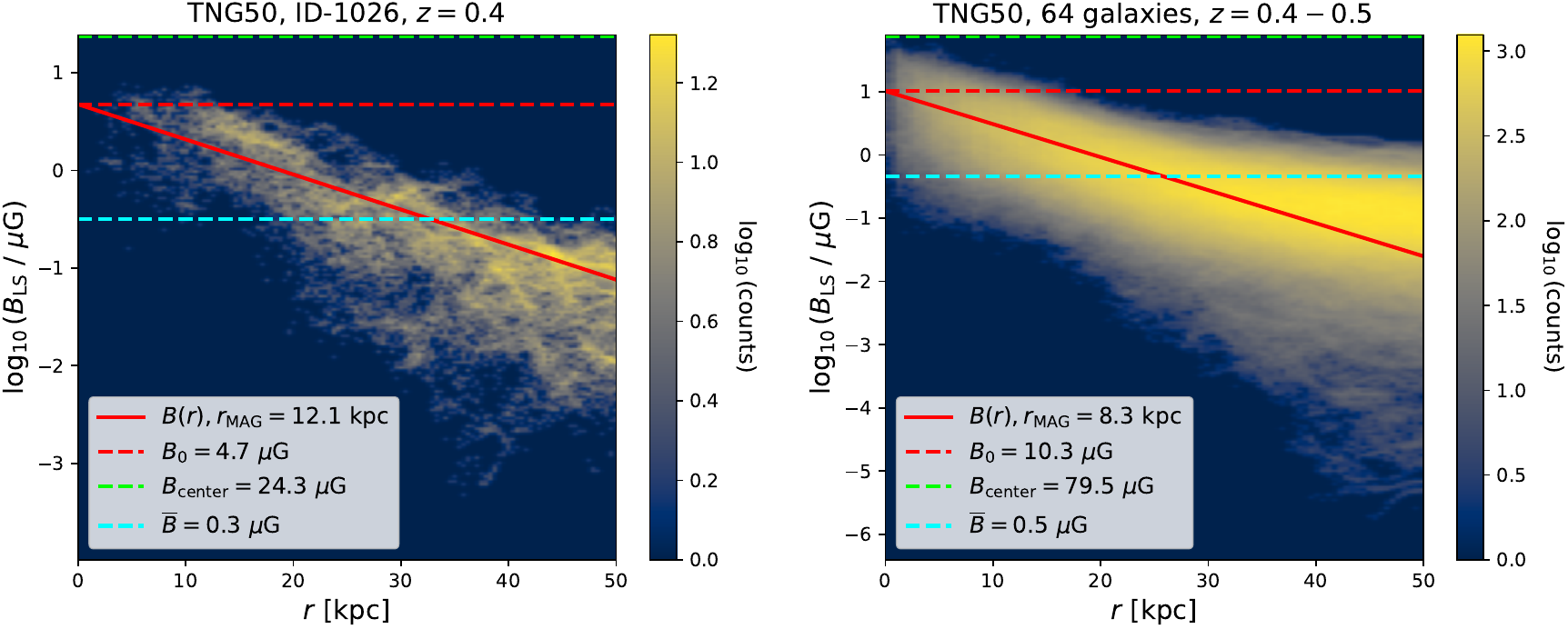}
	\caption{Left panel: For the exemplary galaxy ID-1026 depicted in Fig.\ \ref{fig:InputParams} we plot the radial distribution of the large-scale magnetic field strength within a characteristic disk height $h$. The horizontal dashed lines represent the following magnetic field strength values: $B_{\mathrm{center}}$ at the center of the gas mass (red), the fit parameter $B_{\mathrm{0}}$ (red), and the mean $\overline{B}$ (cyan), respectively. The red solid line indicates the best fit result of the exponential law $B(r)$ (see Eq.~\ref{eq:exp_fit}). Right panel: The same as the left panel, but now for all 64 considered TNG50 galaxies.}
	\label{fig:MagRad}
\end{figure*}

A magnetized galaxy containing cosmic-ray electrons should produce diffuse synchrotron emission. But that is different from saying that such emission has been observationally detected in the lensing galaxy \textsc{B1152+199}. Potentially, such synchrotron emission from the lensing galaxy might impact the RM observations \citep {Reissl2026}. However, the TNG50 simulation does not include cosmic ray (CR) physics. While CR parameters may be roughly estimated \cite[see e.g.][]{Rappaz2022, Rappaz2024}, without radio observations of \textsc{CLASS B1152+199} or similar lensing systems we do not explore the impact of a synchrotron active foreground galaxy within the scope of this paper. The distributions of the input parameters of gas, magnetic field, and thermal electrons, respectively, for one representative TNG50 galaxy are depicted in Fig.~\ref{fig:InputParams}.

We note that  \cite{Mao2017NatAs} introduced the two separate cases of an axisymetric (AS hereafter) as well as a bisymmetric (BS hereafter) magnetic field morphology within the lensing galaxy to model their data. However, as depicted in Fig.~\ref{fig:InputParams} the TNG50 galaxies show in general no clear trend to belong to one of these two categories (see also Fig.~\ref{fig:RM_maps} and \citealt{Kovacs2024}). 

To characterize the large-scale magnetic field independently of the total magnetic field strength, we follow the signed-component projection approach
of \cite{Kovacs2024}. The magnetic field is then decomposed into its cylindrical components, $B_{\mathrm{r}}$, $B_\phi$, and $B_{\mathrm{z}}$. For each component we calculate the signed vertical average through a slab of half-height $h=1\,{\rm kpc}$ characteristic scale height of the disk \citep{Pillepich2019,Kovacs2024}  via
\begin{equation}
    \overline{B}_{\mathrm{j}}(x,y) =
    \frac{1}{2h}
    \int_{-h}^{+h} B_{\mathrm{j}}(x,y,z)\,{\rm d}z,
\end{equation}
with ${ \mathrm{j}\in\{\mathrm{r},\phi,\mathrm{z}\} }$. 
Importantly, the sign of each magnetic-field component is retained during this averaging. We subsequently define the local amplitude of the
vertically averaged large-scale field as
\begin{equation}
    B_{\rm LS}(x,y) =
    \left[
        \overline{B}_r^2(x,y) +
        \overline{B}_\phi^2(x,y) +
        \overline{B}_z^2(x,y)
    \right]^{1/2}.
\label{eq:LS}
\end{equation}
This signed averaging is preferable to averaging the magnetic-field magnitude, $\langle |\boldsymbol{B}| \rangle$, for the present purpose. Small-scale magnetic fluctuations with changing field directions tend to cancel along the vertical integration path, whereas a field component with a persistent direction survives the averaging. In contrast, averaging
$|\boldsymbol{B}|$ removes the sign information and therefore retains both ordered and randomly oriented magnetic-field contributions. The signed projection also preserves the spatial structure in the disk plane, allowing large-scale field reversals to remain visible rather than being cancelled by a galaxy-wide average.

We therefore use $B_{\rm LS}$ as a measure of the large-scale magnetic-field amplitude in the simulations. We stress that this procedure does not
constitute a unique decomposition into coherent and random magnetic fields: fluctuations coherent over the vertical averaging scale can survive the
projection, while reversals along the vertical direction can partially cancel. Nevertheless, it provides an characterization of the large-scale field without imposing an AS or BS magnetic-field geometry on the simulated galaxies.

To facilitate comparison with the observational approach of \citet{Mao2017NatAs}, we fit the exponential profile
\begin{equation}
    B(r) = B_{\mathrm{0}}\ e^{ -r / r_{\mathrm{MAG}}}
    \label{eq:exp_fit}
\end{equation}
to the radial distribution of the large-scale magnetic
field strength $B_{\rm LS}$ in each TNG50 galaxy.
Here, $r$ is the galactocentric cylindrical radius,
$B_0$ is the fitted central amplitude, and
$r_{\mathrm{MAG}}$ is the magnetic field scale length. In Fig.~\ref{fig:MagRad} we show the fit results to the radial distribution of $B(r)$ for the exemplary TNG50 galaxy ID-1026 (left) as well as for the entire set of selected galaxies (right).  

The entire selection of galaxies, as depicted at the right side of Fig.~\ref{fig:MagRad}, has a magnetic field parameter of $B_{\mathrm{0}} = 10.3\ \mu\mathrm{G}$, an average magnetic field within the disks of $\overline{B} = 0.5\ \mu\mathrm{G}$, a peak value of $B_{\mathrm{center}} = 79.5\ \mu\mathrm{G}$ at the center of mass, and a radial scale length of $r_{\mathrm{MAG}}=8.3\ \mathrm{kpc}$. We report that for fitting the magnitude of the magnetic field instead of its filtered large-scale component the fitting parameters are a little larger with $B_{\mathrm{0}} = 12.5\ \mu\mathrm{G}$ and  $r_{\mathrm{MAG}}=12.6\ \mathrm{kpc}$. respectively.

\begin{figure*}[]
	\centering
	\includegraphics[width=1.0\textwidth]{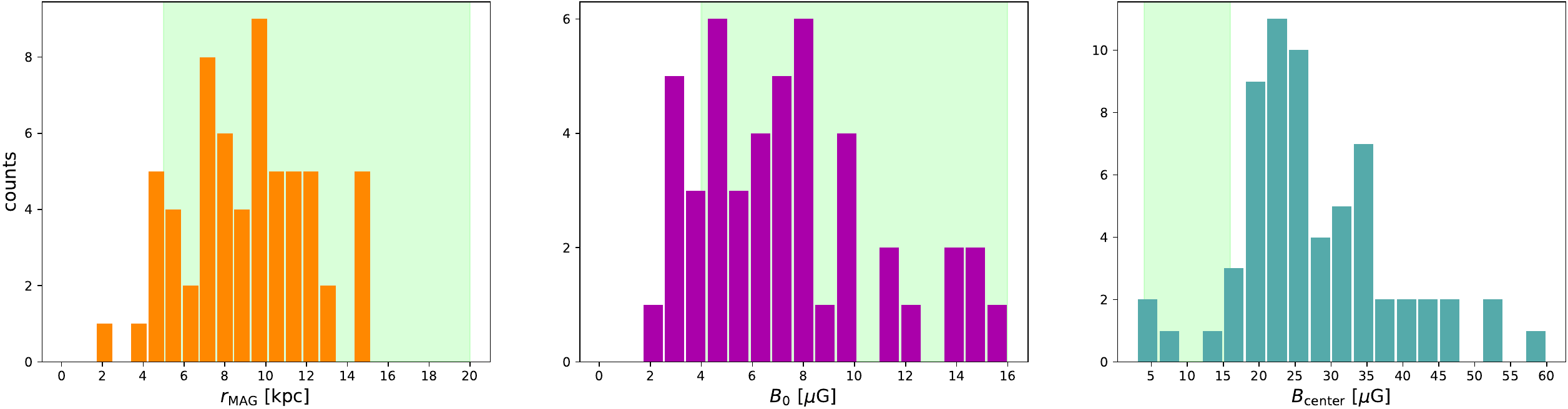}
	\caption{ The distribution of the radial magnetic field scale length $r_{\mathrm{MAG}}$ (left panel), the disk large-scale magnetic field strength $B_{\mathrm{0}}$ (middle panel), and the central magnetic field strength $B_{\mathrm{center}}$ (right panel), respectively, for the selection of considered TNG50 galaxies. Vertical green shaded areas represent the range of values reported in \cite{Mao2017NatAs}.}
	\label{fig:MagHistogramms}
\end{figure*}

In Fig.~\ref{fig:MagHistogramms} we present the distribution of fit parameters for the individual galaxies. We note that the resulting range of 
${ B_{\mathrm{0}} = 1.9\ \mu\mathrm{G} - 15.6\ \mu\mathrm{G} }$
and ${ r_{\mathrm{MAG}}  = 2.2\ \mathrm{kpc} - 15.5\ \mathrm{kpc}  }$, respectively, are broadly compatible with the values reported in
\cite{Mao2017NatAs}. However, given the uncertainties in
the physical properties of the lensing galaxy, this agreement
does not imply that the selected TNG50 galaxies are exact
counterparts of \textsc{CLASS B1152+199}.

\subsection{Synthetic observations} \label{sect:synthobs}

Observationally, magnetic field properties can be probed by multi-wavelengths observations of Faraday rotation, where the plane of linearly polarized radiation along the line-of-sight (LOS) rotates by an angle of ${ \Delta \chi=\chi_{\mathrm{0}} + \phi \lambda^2 }$ when passing through a magnetized and ionized medium. Here, the quantity $\lambda$ is the wavelength, $\chi_{\mathrm{0}}$ is the intrinsic polarization angle of the background source, and the Faraday depth is defined to be
\begin{equation}
\phi=\frac{e^3}{2\pi\ m _{\mathrm{e}}^2c^4}\int_0^{\ell} n _{\mathrm{e}}(\ell^') B _{\mathrm{\mathrm{LOS}}}(\ell^')\  \mathrm{d}\ell^' = k \int_0^{\ell}   n _{\mathrm{e}}(\ell^') B _{\mathrm{\mathrm{LOS}}}(\ell^')\ \mathrm{d}\ell^' \;,
\label{eq:FaradayDepth}
\end{equation}
where the constant $e$ is the electron charge and, subsequently, ${ k=0.81\ \mathrm{rad\ m^{-2} \ pc^{-1}\ \ cm^{-3}}\ \mu \mathrm{G^{-1} } }$ \citep{Burn1966}. The LOS magnetic field strength $B _{\mathrm{\mathrm{LOS}}}$ is modulated by the number density  $n _{\mathrm{e}}$ of free electrons along the path element $\mathrm{d}\ell$ towards the observer, where $\ell=0$ is the position of the polarized source. The Faraday depth $\phi$ is defined to be positive when $B _{\mathrm{\mathrm{LOS}}}$ points in the direction of the observer, and vice verse. Related to the Faraday depth $\phi$ is the rotation measure $\mathrm{RM}=\mathrm{d}  \chi/\mathrm{d}  \lambda^2 $. Ideally, $\phi = \mathrm{RM}$, but effects such as beam depolarization or a synchrotron radiation emitting medium renders this simple identity invalid \cite{Burn1966,Sokoloff1998}. 

We note that in an expanding universe the Faraday depth $\phi$ needs to be corrected by a factor of $(1+z)^{-2}$. However, our RT modeling does neither include the expansion of lengths along the LOS nor any foreground pollution from the Milky Way. Hence, Eq.~\ref{eq:FaradayDepth} represents already the corrected Faraday depths $\phi$ caused by the lensing galaxy in its rest frame. Maps of the integrated Faraday depth $\phi$ for all the  64 selected galaxies are provided in Appendix \ref{app:maps}.

The uncertainties in deriving magnetic field properties by means of Faraday rotation are related to the fact, that neither the Faraday depth $\phi$ nor the rotation measure RM are directly accessible by observations, but instead need to be reconstructed indirectly from multi-wavelengths polarization measurements. Hence, the impact of observational effects need to be explored by constructing  synthetic polarization data. To do so, we perform radiative transfer (RT) calculations using POLARIS\footnote{\url{https://portia.astrophysik.uni-kiel.de/polaris/}}  \citep[][]{Reissl2016,Reissl2019}. The code was originally developed to predict polarized radiation of aligned dust grains \citep{Reissl2016}. We use a version of the code extended to simulate polarized synchrotron emission including Faraday rotation, based on the spatial 3D distribution of gas density, gas temperature, magnetic field, free electrons as well as CRs \citep{Reissl2019}. The advantage of POLARIS is, that it runs in a parallelized mode and solves the RT problem on the native moving mesh Voronoi grid of the  AREPO code \citep[][]{Springel2001}.

Individual rays are traced in the Stokes vector formalism in all four parameters $\vec{S}=\left(I,Q,U,V\right)^T$ simultaneously, where $I$ is the total intensity of radiation, $Q$ and $U$ represent linear polarization and $V$ circular polarization. For each ray the polarized RT problem without CR emission may be written as a equation in matrix form \citep{Rybicki1986,Bohren1998}
\begin{equation}
\vec{S}_{\mathrm{obs}}=\int_0^{\ell} -\hat{K}\vec{S}\ d\ell\, .
\label{eq:RTProblem1}
\end{equation}
Here,  the quantity $\hat{K}$ is the $4\times4$ M{\"u}ller matrix, where the diagonal elements $K_{\mathrm{ii}}$ are the total absorption of intensity $I$. The off-diagonal matrix elements describe the transfer between the Stokes parameters due to differential absorption where the elements of Faraday conversion ${ K_{\mathrm{34}} = -K_{\mathrm{43}} }$ and Faraday rotation ${ K_{\mathrm{23}} = -K_{\mathrm{32}} }$ depend on the local distribution of thermal electrons $n_{\mathrm{th}}$ and the magnetic field $\vec{B}$ during the runtime  (we refer to \citealp{Shcherbakov2008}, \citealp{Dexter2016}, and \citealp{Pandya2016} for further details). Hence, the RT problem in the Stokes vector formalism is connected to the Faraday depth via  ${ \mathrm{d}\phi = \lambda^{-2} K_{\mathrm{32}}\mathrm{d}\ell }$. The code POLARIS solves this equation using a Runge-Kutta-Fehlberg solver (RFK45) with adaptive step size refinement to keep the relative error below $10^{-6}$ \citep[see Appendix A of][]{Reissl2019}.

After solving the RT equation along the entire path, all rays are collected by POLARIS on a detector with $1000 \times 1000$ pixel per wavelength, where the side-length corresponds the extension of the post-processed TNG50 cutout. With the detected Stokes parameters the degree of linear polarization of the radiation $P_{\mathrm{l}}$ and the corresponding orientation angle $\chi$ per wavelength are defined via
\begin{equation}
P_{\mathrm{l}}=\sqrt{\frac{U^2+Q^2}{I^2}}
\label{eq:PlDefinition}
\end{equation}
and
\begin{equation}
\chi=\frac{1}{2} \tan^{-1}\left(\frac{-U}{Q} \right)\, ,
\label{eq:LinearOrientation}
\end{equation}
respectively. 

The observations reported by \cite{Mao2017NatAs} are performed with a resolution of ${ 2.36\arcsec \times 1.75\arcsec }$ at $1.006\ \mathrm{GHz}$ and ${ 0.23\arcsec \times 0.22\arcsec }$ at $7.991\ \mathrm{GHz}$ with a channel width of $1\ \mathrm{MHz}$ for the lower frequency regime up to $2\ \mathrm{MHz}$ in the upper regime averaged into $8\ \mathrm{MHz}$ channels for proper analysis. However, due to the computational cost of such a small resolution in frequency to our RT modeling we apply a larger resolution with 221 channels resulting in a width of $34.8\ \mathrm{MHz}$. For simplicity we utilize a frequency range of exactly $1\ \mathrm{GHz} - 8\ \mathrm{GHz}$ corresponding to a minimal and maximal wavelengths of $\lambda_{\mathrm{min}}=3.75\ \mathrm{cm}$ and 
 $\lambda_{\mathrm{max}}=29.98\ \mathrm{cm}$, respectively.

In a final post-processing step we apply a Gaussian beam with a full width at half maximum (FWHM) of $0\arcsec$, $0.25\arcsec$, $0.5\arcsec$, $1.0\arcsec$, and $2.0\arcsec$, respectively, in order to explore the impact of instrumental effects. The various beam sizes with respect to the spatial extension of the exemplary galaxy ID-1026 are depicted in Fig.~\ref{fig:InputParams}.

\subsection{Source selection} \label{sect:source}

\begin{figure*}[]
	\centering
	\includegraphics[width=0.9\textwidth]{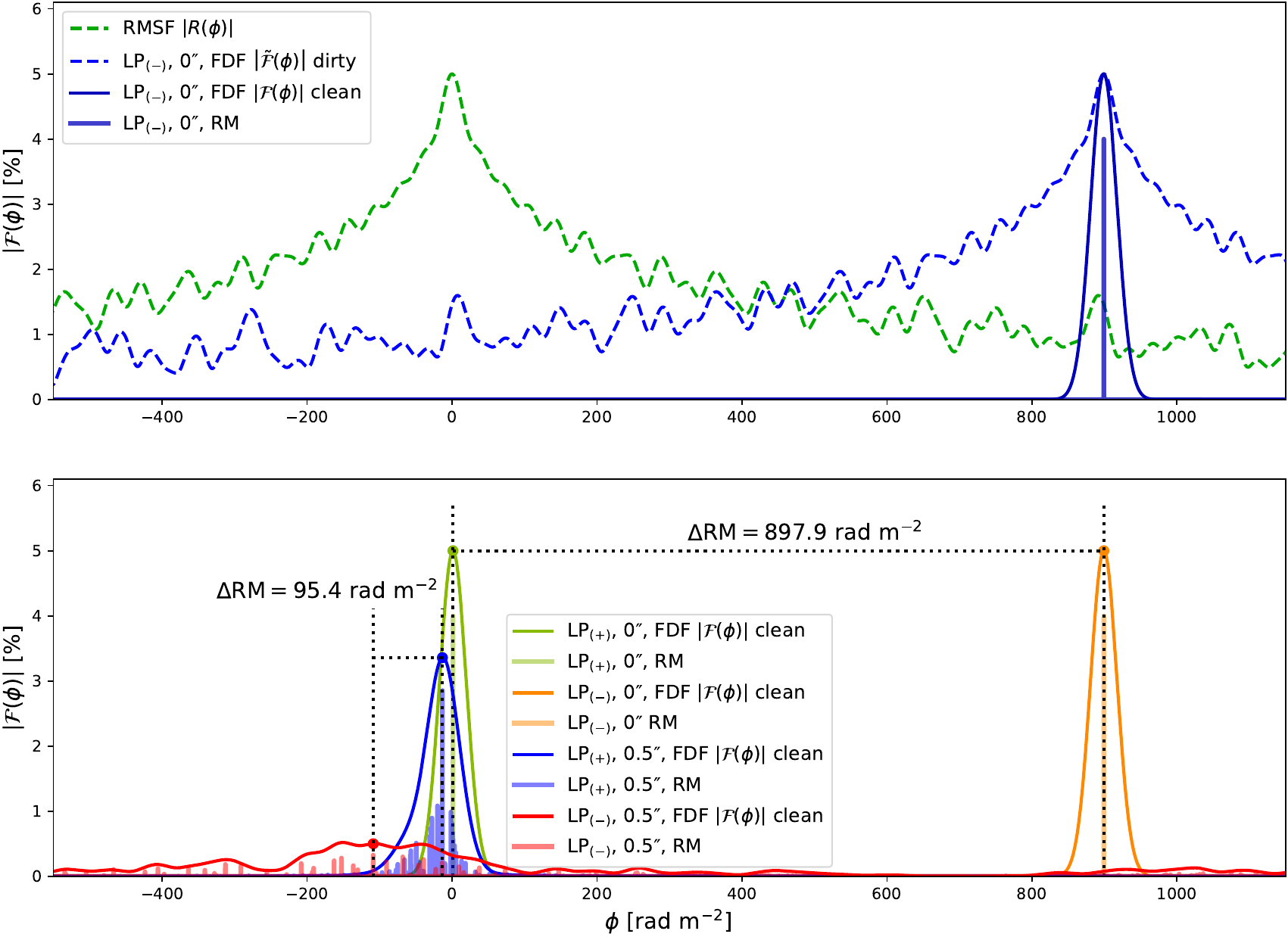}
	\caption{Top panel: The RMSF $R(\phi)$ (dashed green) scaled to the peak of the signal, as well as the observed dirty Faraday spectrum $\tilde{\mathcal{F}}(\phi)$ (dashed dark blue), the de-convolved and cleaned FDF $\mathcal{F}(\phi)$ (solid blue), and the scaled RM component (vertical blue peak) of the negative lensing point LP$_{(-)}$. Bottom panel: The same as the top panel, but for LP$_{(-)}$ as well as LP$_{(+)}$ for a beam size of $0\arcsec$ and $0.5\arcsec$, respectively. For $0\arcsec$, the difference  between the points LP$_{(-)}$ (green) and LP$_{(+)}$ (light blue) leads to $\Delta\mathrm{RM} = 897.9\ \mathrm{rad\ m}^{-2}$, whereas in the case of a beam size of $0.5\arcsec$ we obtain a strikingly different value of $\Delta\mathrm{RM} = 95.4 \ \mathrm{rad\ m}^{-2}$ from the difference between LP$_{(-)}$ (red) and LP$_{(+)}$ (light blue). We note that the considered lensing points LP$_{(-)}$ and LP$_{(+)}$ are identical to those depicted in Fig.~\ref{fig:InputParams}. }
	\label{fig:SpectrumBeam}
\end{figure*}

The initial conditions of the RT problem are defined by the characteristic spectral polarized emission of the background source i.e. ${ \vec{S}(\ell = 0)} = \vec{S}(z=  1.0) = \vec{S}_{\mathrm{source}} $. Since Faraday rotation is independent of intensity we define an ideal background source $\vec{S}_{\mathrm{source}} = (1,\ 0.05,\ 0,\ 0)^T$ that has a polarization and an orientation angle that is constant over frequency. This initial condition corresponds to a degree of linear polarization of $P_{\mathrm{l}}=5\ \%$ and an orientation angle pf $\chi=0^\circ$, independent of wavelength. However, recovering individual RM components within the lensing galaxy heavily depends on the spectral stability of the polarization signal coming from the background source.

In order to explore the impact of variable background sources we take the spectral emission of the two well known calibrator sources 3C48 and 3C286, respectively \citep{Perley2013a,Perley2013b}. The source 3C48 is a compact quasar with a linear polarization ${ P_{\mathrm{l}} = 0.3\% - 5.3\% }$ with an observed  step decline rotation angle of ${ \chi = {-64}^\circ\  \mathrm{to}\ {-150}^\circ }$ resulting in an intrinsic offset of $\mathrm{RM}_{\mathrm{0}} = -72\ \mathrm{rad\ m}^{-2}$ \cite{Perley2013b,Reissl2026} for the observed wavelength regime utilized in \cite{Mao2017NatAs}. The source 3C286 is a quasar with a linear polarization ${ P_{\mathrm{l}} = 8.6\% - 11.9\% }$. In contrast, the polarization angle of 3C286 is nearly constant within the relevant wavelength regime with small variations ${\chi = 33^\circ - 34^\circ }$ and the intrinsic rotation measure of 3C286 is $\mathrm{RM}_{\mathrm{0}} = 0\ \mathrm{rad\ m}^{-2}$. Consequently, 3C286 is better suited to probe the magnetic field strength within the lensing galaxy due to its stable polarized radiation.

As outlined by \cite{Perley2013b} the wavelength dependent Stokes parameters of these sources may be approximated by a cubic polynomial. We use the fit presented by \cite{Reissl2026}. to derive the parameters $I_{\mathrm{source}}$, $Q_{\mathrm{source}}$, and $U_{\mathrm{source}}$, respectively, of $\vec{S}_{\mathrm{source}}$ for both 3C48 and 3C286 assuming a peak in total emission $I_{\mathrm{galaxy}}=1\ \mathrm{a.u.}$ at $1\ \mathrm{GHz}$.

\subsection{Lensing Points} 
\label{sect:lensing}

In order to generate a set of correlated points for the synthetic magnetic field estimations we mimic the lensing of the background source based on the relative position and the mass for each TNG50 galaxy. For simplicity we assume the lensing of the polarized light rays on a point mass. Hence, the mass within the galactocentric radius $\xi$ radius is evaluated via 
\begin{equation}
M(\xi) = 2\pi \int_{0}^{\xi} \xi^' \Sigma_{\mathrm{tot}}(\xi^')  \mathrm{d}\xi^'\;,
\end{equation}
where $\Sigma_{\mathrm{tot}}(\xi^')$ is the total surface density of each TNG50 galaxy considering the masses of the gas, stellar and DM components within $\xi$. In detail, 200 independent positions are randomly sampled homogeneously for the background source. The lensing in the point mass approximation yields two distinct solutions for the observed angle $\theta$ of the apparent position of the lensed source
\begin{equation}
\theta_{(\pm)}=\frac{1}{2}
\left(\beta \pm \sqrt{\beta^2+4\theta_{\mathrm{E}}^2}\right)\, ,
\end{equation}
where the angle $\beta$ is between the optic axis and the true source position. The Einstein radius
\begin{equation}
\theta_{\mathrm{E}}=\sqrt{\frac{4 G M(\xi)}{c^2} \frac{D_{\mathrm{ls}}}{D_{\mathrm{l}}D_{\mathrm{s}}}  }
\end{equation}
describes the ring shaped image for a source positioned exactly on the optical axis of the system. 
The angular size distances between observer and source, $D_{\mathrm{s}}=1675.3\ \mathrm{Mpc}$, observer and lensing galaxy, $D_{\mathrm{l}}=1272.5\ \mathrm{Mpc}$, as well as in between source and galaxy $D_{\mathrm{ls}}=688.6\ \mathrm{Mpc}$ are calculated \citep[for further details see e.g.][]{Peebles1993} using the default cosmological parameters utilized by \cite{Mao2017NatAs}.

The two corresponding lensing points (LP) are labeled $\mathrm{LP}_{(+)}$ and $\mathrm{LP}_{(-)}$ in the following. Two exemplary LPs are depicted in Fig.~\ref{fig:InputParams}, along with their relative positions with respect to the center of the particular TNG50 galaxy ID-1026. We note that the LPs are always collinear, as they are modeled using the point lens approximation. Depending on the lensing model and the assumed parameters of the  system \textsc{CLASS B1152+199}, this is not necessarily the case as outlined in \cite{Zhang2022}. However, a more realistic lensing model, including RT with Faraday rotation, goes beyond the scope of this paper. We estimate that the general statistics presented in the following sections would not differ significantly using a more sophisticated lensing model as we evaluate only the peak intensity of the lensed background source anyway.

\subsection{Faraday rotation measure synthesis}\label{sect:FaradaySynthesis}

Various contributions of Faraday rotation combine together within a single beam, or due to poor resolution in frequency space. Furthermore, multiple source of polarized radiation along the LOS can pollute the observed signal. In these cases, the rotation measure RM and the Faraday depth $\phi$ are no longer identical. 
As in \citet{Mao2017NatAs}, we use Faraday rotation measure (RM) synthesis \citep{Brentjens2005} to analyze the synthetic
observations and recover the Faraday depth $\phi$ as well as individual RM components at the distinct LPs. Here, it is convenient to express the linear polarization as an complex quantity 
\begin{equation}
P_{\mathrm{l}}( \lambda^2 ) = q + \iu u\;,
\label{eq:plin_complex}
\end{equation}
where we utilize the normalized components ${ q=Q/I }$ and ${ u=U/I }$, respectively, expressed in percent. Essential for RM synthesis is the Faraday spectrum or Faraday dispersion function (FDF) 
\begin{equation}
\mathcal{F}( \phi ) = \int_{-\infty}^{\ \infty} P( \lambda^2 )  \exp\left( -2\pi \iu \phi \lambda^2 \right) \mathrm{d} \lambda^2\;,
\label{eq:fdf}
\end{equation}
which represents the Fourier transform of the linear polarization $P_{\mathrm{l}}$ at Faraday depth $\phi$. Naturally, any source can only be observed within a limited range of ${ \lambda \in \{\lambda_{\mathrm{min}},\lambda_{\mathrm{max}}\} }$. This limitation is taken into account via the weight function
\begin{equation}
\mathcal{W}( \lambda^2 ) =\begin{cases} 1  &\mbox{if } \lambda_{\mathrm{min}}^2 \leq \lambda^2 \leq \lambda_{\mathrm{max}}^2 \\ 
		    0  & \mbox{otherwise} \end{cases}\;.
\end{equation}
Consequently, the dirty FDF is
\begin{equation}
\tilde{\mathcal{F}}( \phi ) = \int_{-\infty}^{\ \infty} P( \lambda^2 ) \mathcal{W}( \lambda^2 )  \exp\left( -2\pi \iu \phi \lambda^2 \right) \mathrm{d} \lambda^2\propto ( P\ *\ R) ( \phi )
\label{eq:m}
\end{equation}
is a function different from the clean FDF $\mathcal{F}( \phi )$. Here, $ { (P*R)  }$ is the  convolution of clean FDF and the rotation measure transfer function \footnote{Rotation Measure Spread Function (RMSF) in the original paper \citep{Burn1966} on RM synthesis.} (RMTF) $R ( \phi )$ is the point spread function in Faraday depth space and is defined by the Fourier transform of the weight function. In Fig.~\ref{fig:SpectrumBeam} we depict the RMTF $R( \phi )$ for the particular wavelengths range utilized in this paper for the RT calculation of synthetic observation.

To identify individual RM components at the LPs we use the  RM CLEAN algorithm. The algorithm attempts to separate the contribution of RMSF $R(\phi)$ from the dirty FDF $\tilde{\mathcal{F}}(\phi)$ and ,subsequently,  reconstruct the FDF $\mathcal{F}(\phi)$ by regarding it as a collection of delta functions \citep{Hogbom1974,Brentjens2005,Heald2009}. The synthesized observation are cleaned with the help of the RMTools package\footnote{\url{https://github.com/CIRADA-Tools/RM-Tools}}\citep[][]{Purcell2020} modified to handle the POLARIS output data directly. The dirty FDF as well as the steps of the cleaning process are depicted in Fig.~\ref{fig:SpectrumBeam}.  Here, we show that without considering instrumental effects or pollution from the lensing galaxy the FDF $\mathcal{F}(\phi)$ is correlated to a single RM component. We emphasize that the height of the maximum of $\mathcal{F}(\phi)$ recovers the $P_{\mathrm{l}}=5\ \%$ assumed for the background source. For simplicity, we identify the location of the peak of the clean FDF as the dominant RM component at the distinct LPs of the lensing galaxy even though multiple components might be present in the Faraday spectrum. This simplification is consistent with the method outlined in \cite{Mao2017NatAs}.

\begin{figure}[]
	\centering
	\includegraphics[width=0.49\textwidth]{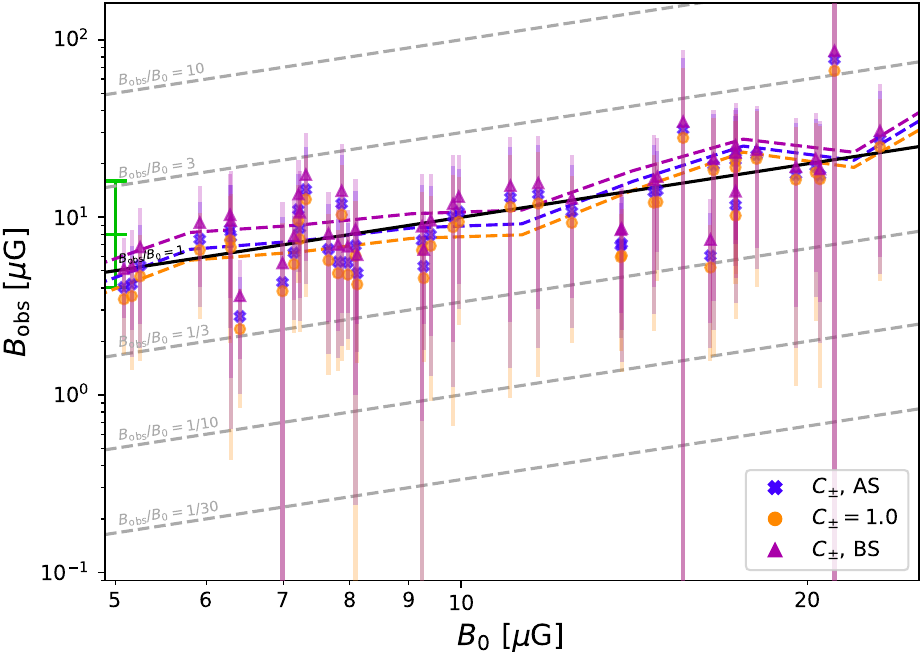}
	\caption{The magnetic field $B_{\mathrm{obs}}$ derived from observable parameters (compare Eq.~\ref{eq:mag_obs}) for a beam size of $0\arcsec$ versus the best-fit parameter $B_{\mathrm{0}}$ characterizing the large-scale field in the simulations. Error bars present a range of one standard deviation. We consider three different scenarios where the geometrical factors $C_\pm$ are calculated for axis-symmetric (AS) field (blue crosses), are constant $C_\pm = 1$ (yellow dots),  and being calculated for a bisymetrical (BS) field (purple triangles). Dashed lines indicate the running mean. The black solid line represent the one-to-one correlation between $B_{\mathrm{obs}}$ and $B_{\mathrm{0}}$, where as dashed gray lines indicate deviations by a factor of three and ten, respectively. The range of $B_{\mathrm{obs}}$ indicated by vertical green lines show the observed values for the axis-symmetric field model \citep{Mao2017NatAs}.}
	\label{fig:MagGeometry}
\end{figure}

\begin{figure}[]
	\centering
	\includegraphics[width=0.49\textwidth]{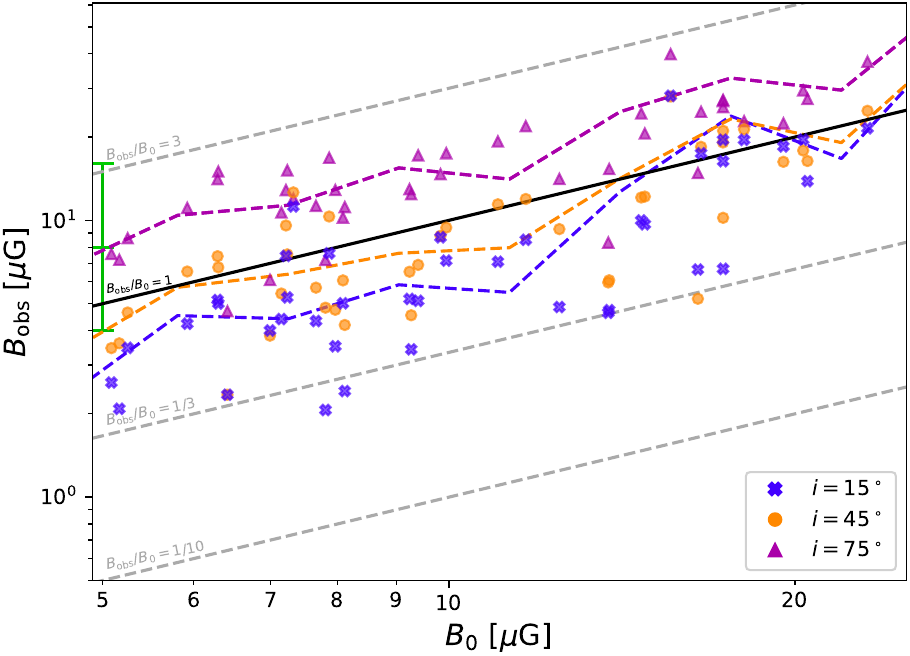}
	\caption{ The same as Fig.~\ref{fig:MagGeometry} but for the three different inclinations angles $i=15^\circ$ (blue crosses), $i=45^\circ$ (yellow dots), and $i=75^\circ$ (purple triangles), respectively. }
	\label{fig:MagInc}
\end{figure}

\begin{figure}[]
	\centering
	\includegraphics[width=0.49\textwidth]{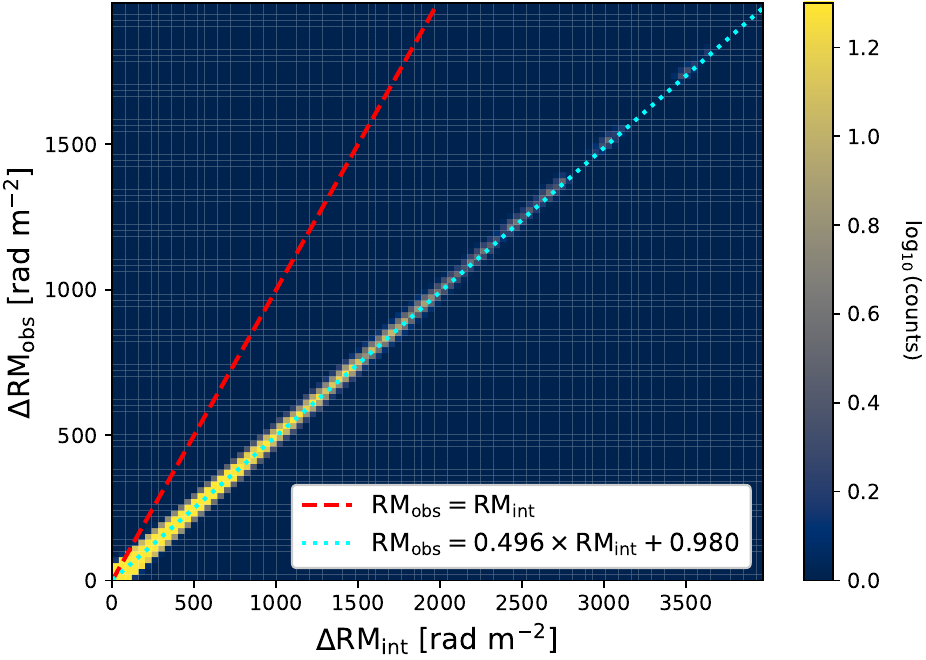}
	\caption{The correlation between the differences in rotation measure $\Delta\ \mathrm{RM}_{\mathrm{int}}$ and $\Delta\ \mathrm{RM}_{\mathrm{obs}}$, respectively. The difference $\Delta\ \mathrm{RM}_{\mathrm{int}}$ is calculated by direct integration through the simulations grid while $\Delta\ \mathrm{RM}_{\mathrm{obs}}$ is derived indirectly by means of RM synthetics for a beam size of $0.25\arcsec$ and an inclination of $i=45^\circ$.The red dashed line indicates the one-to-one correlation between $\Delta\ \mathrm{RM}_{\mathrm{int}}$ and $\Delta\ \mathrm{RM}_{\mathrm{obs}}$ while the cyan line represents the best linear fit between both quantities. We note that the depicted correlation includes the full set of lensing points of all TNG50 galaxies. }
	\label{fig:RMCorr}
\end{figure}

\begin{figure}[]
	\centering
	\includegraphics[width=0.49\textwidth]{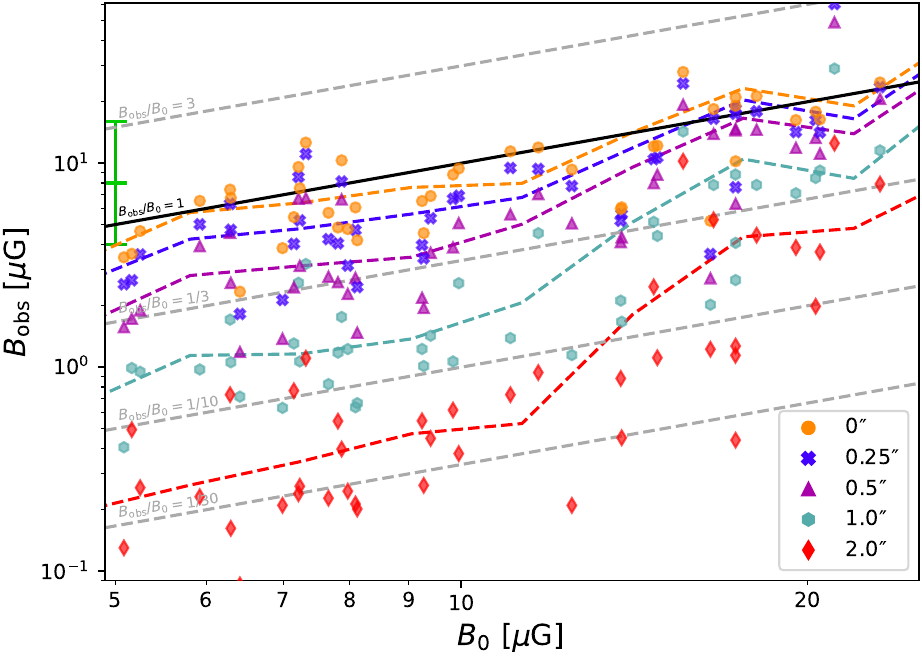}
	\caption{The same as Fig.~\ref{fig:MagInc}, for an inclination of $i=45^\circ$, now showing the impact of beam size on the synthesized data. For the beam we consider values of $0\arcsec$ (yellow), $0.25\arcsec$ (blue), $0.5\arcsec$ (purple), $1.0\arcsec$ (green), and $2.0\arcsec$ (red), respectively.}
	\label{fig:MagBeam}
\end{figure}

\begin{figure*}[]
	\centering
	\includegraphics[width=0.9\textwidth]{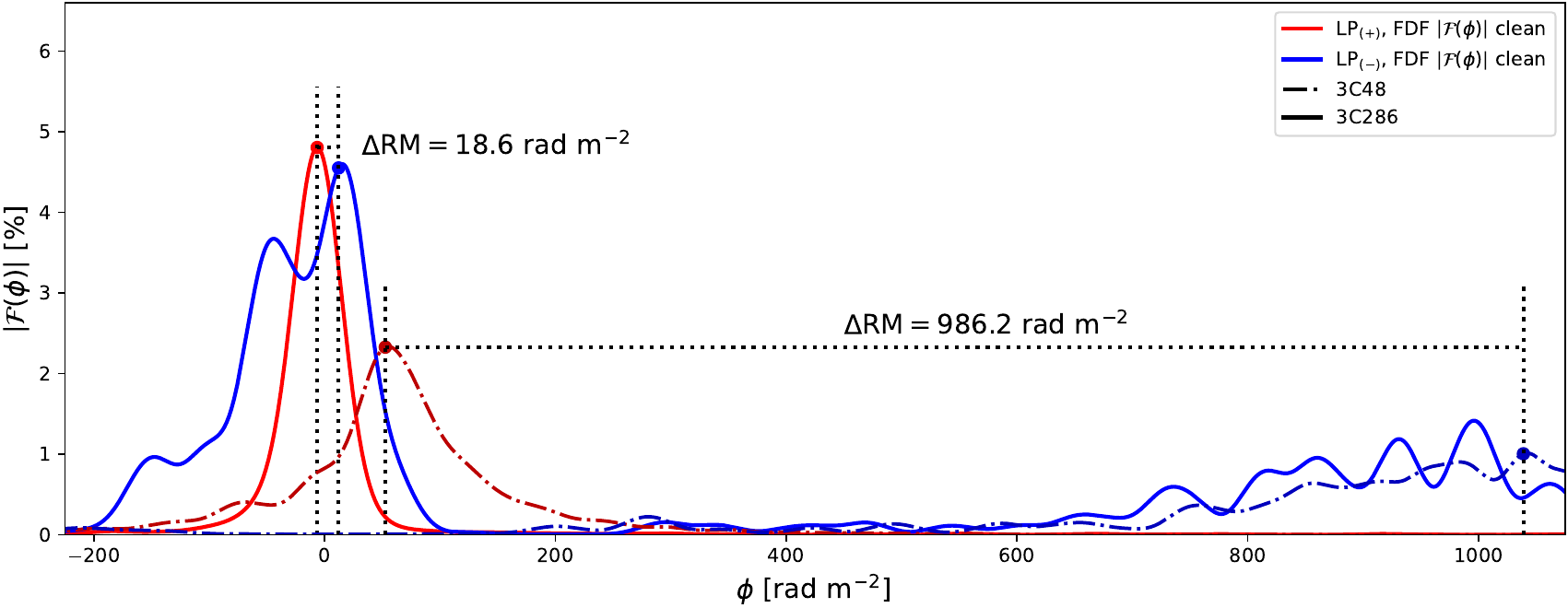}
	\caption{Faraday rotation derived for the TNG50 galaxy ID-1026 (compare  Figs.\ \ref{fig:InputParams} and \ref{fig:SpectrumBeam}). We plot the clean FDF $\mathcal{F}(\phi)$ for a beam size of $0.25\arcsec$, and compare two different contributions from two distinct source modeled based on the well-characterized quasars 3C48 and 3C286. Red and blue lines represent the FDFs at the points LP$_{(+)}$ and LP$_{(-)}$, respectively. Dashed lines correspond to the source 3C48, while solid lines depict results for the source 3C286. The peak value of each FDF is again indicated by a dot. }
	\label{fig:SpectrumCR}
\end{figure*}

\subsection{Magnetic field estimation} \label{sect:mag_est}
In order to estimate the magnetic field strength from synthetic
observations, we express the Faraday rotation measure in terms
of the thermal electron column density $N_{\mathrm{th}}$ and
an effective LOS magnetic field strength $B_{\mathrm{LOS,eff}}$,
such that
\begin{equation}
   \mathrm{RM}
   = k \int n_{\mathrm{th}}(\ell) B_{\mathrm{LOS}}(\ell)
   \ \mathrm{d}\ell
   = k N_{\mathrm{th}} B_{\mathrm{LOS,eff}} \;,
   \label{eq:effective_BLOS}
\end{equation}
where
\begin{equation}
   B_{\mathrm{LOS,eff}}
   =
   \frac{
   \int n_{\mathrm{th}}(\ell) B_{\mathrm{LOS}}(\ell)
   \ \mathrm{d}\ell
   }{
   \int n_{\mathrm{th}}(\ell)\ \mathrm{d}\ell
   } \;.
\end{equation}
This definition accounts for the spatial correlation between
the thermal electron density and the LOS magnetic field,
including possible field reversals along the LOS.
Following the observational approach of \citet{Mao2017NatAs},
we approximate the effective LOS field using an exponential
radial profile and geometrical projection factors.
This introduces a simplified description of the magnetic
field distribution, which we test against the synthetic
observations of the TNG50 galaxies.

Hence, the difference in RM between the two correlated
lensing points LP$_{(+)}$ and LP$_{(-)}$ may be estimated as

\begin{equation}
\begin{split}
   \Delta \mathrm{RM} \approx
   \eta k B_0 \sin i \bigg[
   & C_{(+)} N_{\mathrm{th}(+)}
   \exp\left(-\frac{r_{(+)}}{r_{\mathrm{MAG}}}\right)
   \\[-2pt]
   {}-{}&
   C_{(-)} N_{\mathrm{th}(-)}
   \exp\left(-\frac{r_{(-)}}{r_{\mathrm{MAG}}}\right)
   \bigg] \;,
\end{split}
\end{equation}
where  $\eta=3.24\times10^{-19}\ \mathrm{pc\,cm^{-1}}$ is a unit-conversion factor \citep{Tahani2018},  $k$ is obtained from Eq.~\ref{eq:FaradayDepth}, $\sin i$ accounts for the inclination $i$ of the galaxy, the corrections, $C_{(\pm)}$ are geometrical factors assuming an AS or BS magnetic field morphology \citep[for details see][]{Mao2017NatAs}, $r_{(\pm)}$ are the galactocentric radii in the
$x$-$y$ plane of the synthetic images,
$N_{\mathrm{th}(\pm)}$ are the thermal electron column densities at the corresponding lensing points, and $C_{(\pm)}$ are the geometrical projection factors. The parameters $B_0$ and $r_{\mathrm{MAG}}$ describe the exponential magnetic field profile (see Eq.~\ref{eq:exp_fit}). Finally, the large-scale coherent magnetic field strength within each TNG50 galaxy may be estimated as
\begin{equation}
B_{\mathrm{obs}} \approx \frac{|\Delta \mathrm{RM}|}{\eta k \sin i \left|  C_{(+)} N_{\mathrm{th(+)}}e ^{\frac{-r_{(+)}}{r_{\mathrm{MAG}}}}  - C_{(-)} N_{\mathrm{th(-)}}e ^{\frac{-r_{(-)}}{r_{\mathrm{MAG}}}}  \right|}
\label{eq:mag_obs}
\end{equation}

In contrast to the Faraday rotation measures obtained from our synthetic observations, the thermal electron column densities $N_{\mathrm{th}(\pm)}$ at the corresponding LPs are taken directly from the TNG50 column density maps (see Fig.~\ref{fig:InputParams}). Observationally, thermal electron columns can be constrained through pulsar dispersion measures \citep{Manchester1981,Taylor1993} or inferred from H$\alpha$ emission with additional assumptions about the ionized gas distribution \citep{Reynolds1977,Gutierrez2010}. Since reproducing these independent observational diagnostics is beyond the scope of our polarized RT calculations, we use the simulation-derived column densities and assign them an assumed observational uncertainty. To account for potential instrumental effects, we additionally introduce a white noise term of $2\times10^{-4}$ relative to the intensity $I$ in our synthetic observations. Finally, we assume that the parameters required in Eq.~\ref{eq:mag_obs} are known within the following uncertainty ranges $N_{\mathrm{th}}\pm10\ \%$, $r_{\mathrm{MAG}}\pm2\ \mathrm{kpc}$, and $i\pm 15\ \%$ \citep{Toft2000,Dai2009,Mao2017NatAs}.

\section{Results}
\subsection{Role of magnetic field geometry}
In Fig.~\ref{fig:MagGeometry}, we show the correlation between the fit parameter $B_{\mathrm{0}}$ of the large-scale field and the magnetic field strength $B_{\mathrm{obs}}$ derived from synthetic observations and subsequent RM synthesis for all 64 TNG50 galaxies. Each data point in Fig~\ref{fig:MagInc} represents the average and median values, respectively, of $B_{\mathrm{obs}}$ over multiple realizations within the error range introduced in Sect.\ref{sect:mag_est}. The quantity $B_{\mathrm{obs}}$ is derived using correction factors $C_{\pm}$ (compare Eq.~\ref{eq:mag_obs}) assuming AS and BS field geometries, as well as constant values. We find that the AS and BS field assumptions result in overall slightly higher values of $B_{\mathrm{obs}}$ than the constant case. However, the distribution of the average $B_{\mathrm{obs}}$ is very narrow for the different assumed field morphologies, with all values lying well within the individual error bars. Hence, we conclude that the choice of the geometrical correction factors $C_{\pm}$ has only a minor impact on the statistical distribution of $B_{\mathrm{obs}}$ for our sample. In what follows, it is our aim to isolate the effects of inclination, beam depolarization, and source properties rather than to impose a specific large-scale magnetic field morphology on the TNG50 galaxies. For this we adopt $C_{\pm}=1$ as a reference case.

\subsection{Impact of galaxy inclination} 
\label{sec:inclination}

In Fig.~\ref{fig:MagInc}, we present the results of RM synthesis for different inclination angles of the disk, assuming no beam depolarization. In this scenario, the polarized emission originates only from the ideal background source and is Faraday rotated within the disk of the lensing galaxy.

For an inclination angle of $i = 75^\circ$ (near edge-on orientation), the large-scale toroidal component of the magnetic field in the galaxy’s disk is predominantly parallel to the line of sight (LOS). As a result, the difference $\Delta$RM is nearly maximized for all combinations of LPs. Therefore, the ratio of the averages of the fit parameter $B_{\mathrm{0}}$ and the synthesized field strengths $B_{\mathrm{obs}}$ is ${B_{\mathrm{obs}} / B_{\mathrm{0}} = 1 - 3}$, with a probability of $83.3\ \%$ that the ratio falls within this range. We emphasize that $B_{\mathrm{obs}}$ for individual galaxies may vary considerably within the observed factor of three, but the overall trend remains similar even when assuming a BS or AS magnetic field geometry.

Decreasing the inclination down to $i = 45^\circ$ decreases the difference $\Delta$RM. Consequently, most of the average values of $B_{\mathrm{obs}}$ move closer to or even above the one-to-one line. However, the bulk remains below this line, but above a value of $1/3$. The probability of finding a ratio of $B_{\mathrm{obs}} / B_{\mathrm{0}}$ within the interval ${1/3 - 3}$ is $86.3\ \%$. The general trend of $B_{\mathrm{obs}}$ over $B_{\mathrm{0}}$ becomes slightly steeper, with galaxies having $B_{\mathrm{0}} \approx 10\ \mu\mathrm{G}$ underestimating $B_{\mathrm{obs}}$, and vice versa. For galaxies with $B_{\mathrm{0}} \lessapprox 12\ \mu\mathrm{G}$, most of the TNG50 galaxies fall within the range of ${ B_{\mathrm{obs}}/B_0 = 1/3 - 1 }$, which is consistent with the magnetic field strengths reported for the lensing galaxy in \cite{Mao2017NatAs}, for an AS magnetic field morphology. Here, the inclination is $i = 33^\circ$, and given the ideal conditions, the lensing system \textsc{CLASS B1152+199} should also fall within the range of ratios $B_{\mathrm{obs}} / B_{\mathrm{0}} = {1/3 - 1}$ as well.

For an inclination of $i = 15^\circ$ (near face-on orientation), the average ratio $B_{\mathrm{obs}} / B_{\mathrm{0}}$ is mostly below unity, with a smaller spread of data points compared to larger inclinations. Here, the probability of finding a ratio of $B_{\mathrm{obs}} / B_{\mathrm{0}}$ within ${1/3 - 1}$ is $83.5\ \%$. 

Although error bars are not explicitly shown in Fig.~\ref{fig:MagInc} for clarity, we report that range of errors for each galaxy completely overlaps with the one-to-one relation within one standard deviation (compare Fig.~\ref{fig:MagGeometry}).

Overall, we note that the synthesized average values of $B_{\mathrm{obs}}$ correlate well with the fit parameter $B_{\mathrm{0}}$ for each TNG50 galaxy within a factor of three, even for smaller inclinations. However, the results presented in Fig.~\ref{fig:MagInc} do not include potential depolarization effects and a potentially variable background source that could affect the Faraday rotation signal. These cases we discuss next.

\subsection{Beam depolarization} \label{sec:depol}

In this section, we present results using RM synthesis with various beam sizes. It is important to note that the data presented by \cite{Mao2017NatAs} were observed with resolution dependent on the frequency, with higher resolution at higher frequencies (see Sect.~\ref{sect:synthobs}). For simplicity, we use the same beam size across all frequencies.

Although highly unlikely, since lensing does not rotate the plane of linear polarization \citep{Dyer1992}, we assume that rays along the lensed paths fan out and blend into the entire beam. This scenario represents the worst-case situation, as multiple RM components from different regions of the lensing galaxy’s disk would contribute to the Faraday rotation within each beam, potentially canceling each other out. Consequently, this effect reduces the accuracy of recovering the magnetic field strength $B_{\mathrm{0}}$ considerably.

In Fig.\ref{fig:SpectrumBeam}, we show the impact of beam depolarization on the Faraday spectrum. Without beam depolarization, the exemplary LPs (compare Fig.\ \ref{fig:InputParams}) show a difference of $\Delta\ \mathrm{RM} = 897.9\ \mathrm{rad\ m}^{-2}$, where the Faraday depth distribution function (FDF) represents a single RM component. For a beam size of $0.5\arcsec$ and a light bundle diverging within the galaxy, the central peak of the clean FDF near $\phi = 0.0\ \mathrm{rad\ m}^{-2}$ becomes depolarized from $P_{\mathrm{l}} = 5\ \%$ to $P_{\mathrm{l}} = 3.4\ \%$ and shifts slightly into the negative range of $\phi$. The FDF now contains multiple RM components, but a dominant one can still be identified. The peak at $\phi = 897.9\ \mathrm{rad\ m}^{-2}$ becomes more severely depolarized from $P_{\mathrm{l}} = 5\ \%$ to $P_{\mathrm{l}} = 0.45\ \%$ with a much wider spread in Faraday depth. In contrast to the central peak, a dominant RM component is no longer present within the spectrum, and consequently, the difference declines to $\Delta\ \mathrm{RM} = 95.4\ \mathrm{rad\ m}^{-2}$ for a beam size of $0.5\arcsec$.
However, we need to note that the beam depolarization observed in this particular pair of LPs is not fully representative, as the impact on $\Delta$RM is not as severe in many cases.

In Fig.~\ref{fig:RMCorr} we show the correlation between the difference $\mathrm{RM}_{\mathrm{int}}$, directly derived by integrating though the TNG galaxies along each lensing pair (see Eq.~\ref{eq:FaradayDepth}) and the difference $\mathrm{RM}_{\mathrm{obs}}$ indirectly calculated using RM synthesis (see Sect.~\ref{sect:FaradaySynthesis} and Fig.\ref{fig:RM_maps}). Here, we utilize a beam size of $0.25\arcsec$ and an inclination of $i=45^\circ$. The correlation shows a strictly linear relation with a slight larger spread towards smaller $\mathrm{RM}_{\mathrm{int}}$. However, the trend we report is not a one-to-one correlation but has a smaller slope of $0.496$. Hence, the RM synthesis keeps a clear correlation between observed and actual difference in RM for a pair of lensing points. Any deviations between indirectly derived $B_{\mathrm{obs}}$ the  large-scale field is more likely related to the large uncertainty in the assumed free electrons $N_{\mathrm{th}}$ and the fit parameters $B_{\mathrm{0}}$ and $r_{\mathrm{MAG}}$, respectively. 

 In Fig.~\ref{fig:MagBeam}, we show the maximal impact of beam depolarization for the entire ensemble of TNG50 lensing galaxies, assuming an inclination of $i = 45^\circ$. A beam size of $0.5\arcsec$ generally decreases the observed magnetic field strengths $B_{\mathrm{obs}}$ by a factor of 0.5, while the overall trend, dependent on the fit parameter $B_{\mathrm{0}}$, shows higher values of $B_{\mathrm{obs}}$ for increased values of $B_{\mathrm{0}}$. The probability of finding a data point within the ratio $B_{\mathrm{obs}} / B_{\mathrm{0}} = 1/3 - 1$ is between $47.7\ \%$ and $75.2\ \%$ for beams ranging from $0\arcsec$ to $0.5\arcsec$. As the beam size increases, the Faraday rotation within the beam causes further depolarization, leading to a trend toward smaller $\Delta\ \mathrm{RM}$. For a beam size of $1\arcsec$, we report a probability of $64.0\ \%$ to find data points within the ratio ${1/10 - 1/3}$. In the case of multiple RM components within a $2.0\arcsec$ beam, the recovered values of $B_{\mathrm{obs}}$ per galaxy and over all 200 LPs are, on average, close to the ratio $1/10 - 1/30$, with a probability of $44.9\ \%$ to find a data point in that range.

As mentioned earlier, a scenario in which multiple rays contribute over the entire beam size of $2.0\arcsec$ is highly unlikely. In this extreme case, the magnetic field strength $B_{\mathrm{obs}}$ recovered for each of the TNG50 galaxies is too low by a factor of approximately $1/30$ when compared to its expected value $B_{\mathrm{0}}$. However, it is much more likely that the LPs represent only a single ray from the background source. In this case the quantity $B_{\mathrm{obs}}$ can be considered an more accurate representation of the characteristic magnetic field strength $B_{\mathrm{0}}$ within a factor of three. So far, this conclusion holds for a synchrotron-quiet lensing galaxy. 

\subsection{Frequency dependent polarization of background sources} \label{sec:sync}

\begin{figure}[]
	\centering
	\includegraphics[width=0.49\textwidth]{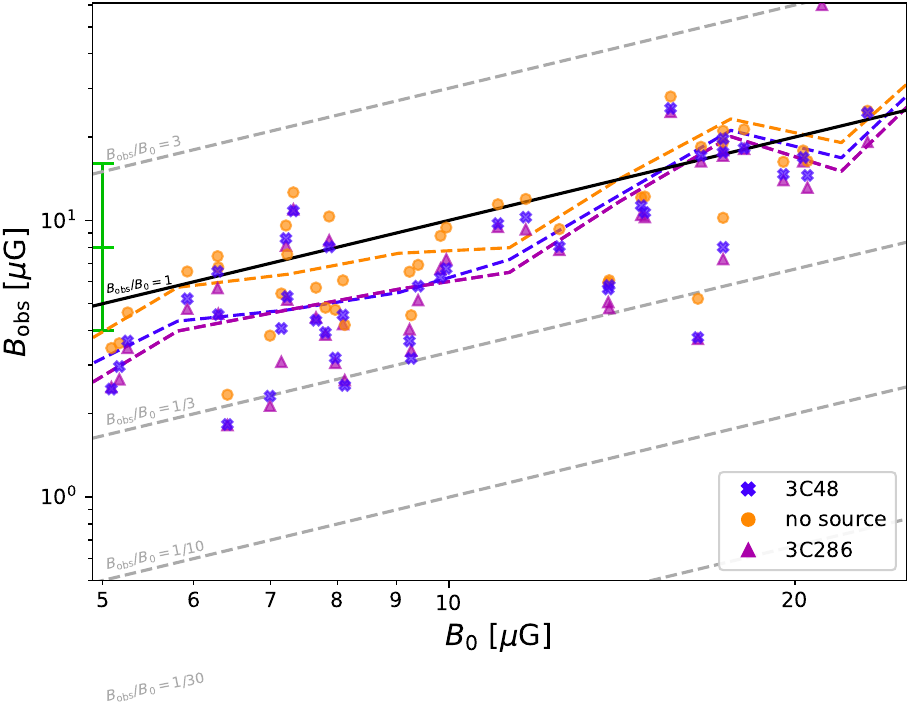}
	\caption{The same as Fig.~\ref{fig:MagBeam} but for different background source an inclination of $i=45^\circ$ of the foreground galaxies and a beam size of $0.25\arcsec$. The source models 3C48 and 3C286 are depicted as blue crosses and purple triangles while the reference model with no background source is depicted by yellow dots.}
	\label{fig:CR}
\end{figure}

The lensing galaxy in the system \textsc{CLASS B1152+199} is a late-type disk galaxy with ongoing star formation \citep{Myers1999, Toft2000, Rusin2002, Dai2009}. The observed coherence of the polarization angle $\chi$ and the degree of polarization $P_{\mathrm{l}}$ strongly suggest that the polarized radiation in the system \textsc{CLASS B1152+199} is dominated by the background source \citep{Mao2017NatAs}. 



In this section, we explore the impact of highly variable polarization properties over frequency of the background sources. Specifically, we investigate Stokes parameters based on the emission of the well-known sources 3C286 and 3C48, respectively. The former source exhibits marginal depolarization and a nearly constant polarization angle $\chi$ over the considered frequency range, while 3C48 shows significant variation in $\chi$.

In Fig.~\ref{fig:SpectrumCR}, we show the Faraday spectra
for two exemplary lensing points (LPs) using the polarized
background sources 3C286 and 3C48 introduced in
Sect.~\ref{sect:source}. For 3C286, the FDF peaks associated
with $\mathrm{LP}_{(+)}$ and $\mathrm{LP}_{(-)}$ occur at
approximately $\phi=-7\ \mathrm{rad\ m^{-2}}$ and
$\phi=12\ \mathrm{rad\ m^{-2}}$, respectively, resulting in
$\Delta\mathrm{RM}=18.6\ \mathrm{rad\ m^{-2}}$.
In contrast, for 3C48 the corresponding FDF peaks occur at
approximately $\phi=47\ \mathrm{rad\ m^{-2}}$ and
$\phi=1033\ \mathrm{rad\ m^{-2}}$, yielding
$\Delta\mathrm{RM}=986.2\ \mathrm{rad\ m^{-2}}$.
These differences illustrate the sensitivity of the recovered RM to the frequency-dependent polarization properties of the background source.

Altogether, we find $\Delta \mathrm{RM} = 986.2\ \mathrm{rad\ m}^{-2}$ for 3C48, and $\Delta \mathrm{RM} = 18.6\ \mathrm{rad\ m}^{-2}$ for 3C286. Again, we note that the LPs in Fig.~\ref{fig:SpectrumCR} are not fully representative, because the depolarization and variations in $\Delta$RM are not as extreme for most of the TNG50 galaxies and LPs, respectively.

In Fig.~\ref{fig:CR}, we present the impact of a source with a frequency dependent polarization for the entire ensemble of selected TNG50 galaxies. For  most of the synthesized magnetic field strengths $B_{\mathrm{obs}}$ are within the range of ${ B_{\mathrm{obs}} = 4\ \mu\mathrm{G} - 16\ \mu\mathrm{G} }$ reported in \cite{Mao2017NatAs}. The strength $B_{\mathrm{obs}}$ synthesized with a beam of $0.25\arcsec$, utilizing the polarized radiation of the sources 3C48 and 3C286, shows a marginally wider spread compared to the ideal case, where the source emits frequency-independent polarized radiation. In this scenario, the variations of The running means show only small variations, with
$72.1\%-84.8\%$ of the inferred values lying within $B_{\mathrm{obs}}/B_0=1/3-1$.

the average $B_{\mathrm{obs}}$ recovered with the two variable background sources, is well within one standard deviation associated with each data point. In conclusion, the field strength $B_{\mathrm{obs}}$ characteristic for each lensing galaxy can mostly likely be recovered within a factor of ${1/3 - 1}$, even if the background source in the lensing system has frequency dependent variations in polarization.

\subsection{Distribution of the difference in RM} \label{sec:distribution}



\begin{figure}[]
	\centering
	\includegraphics[width=0.49\textwidth]{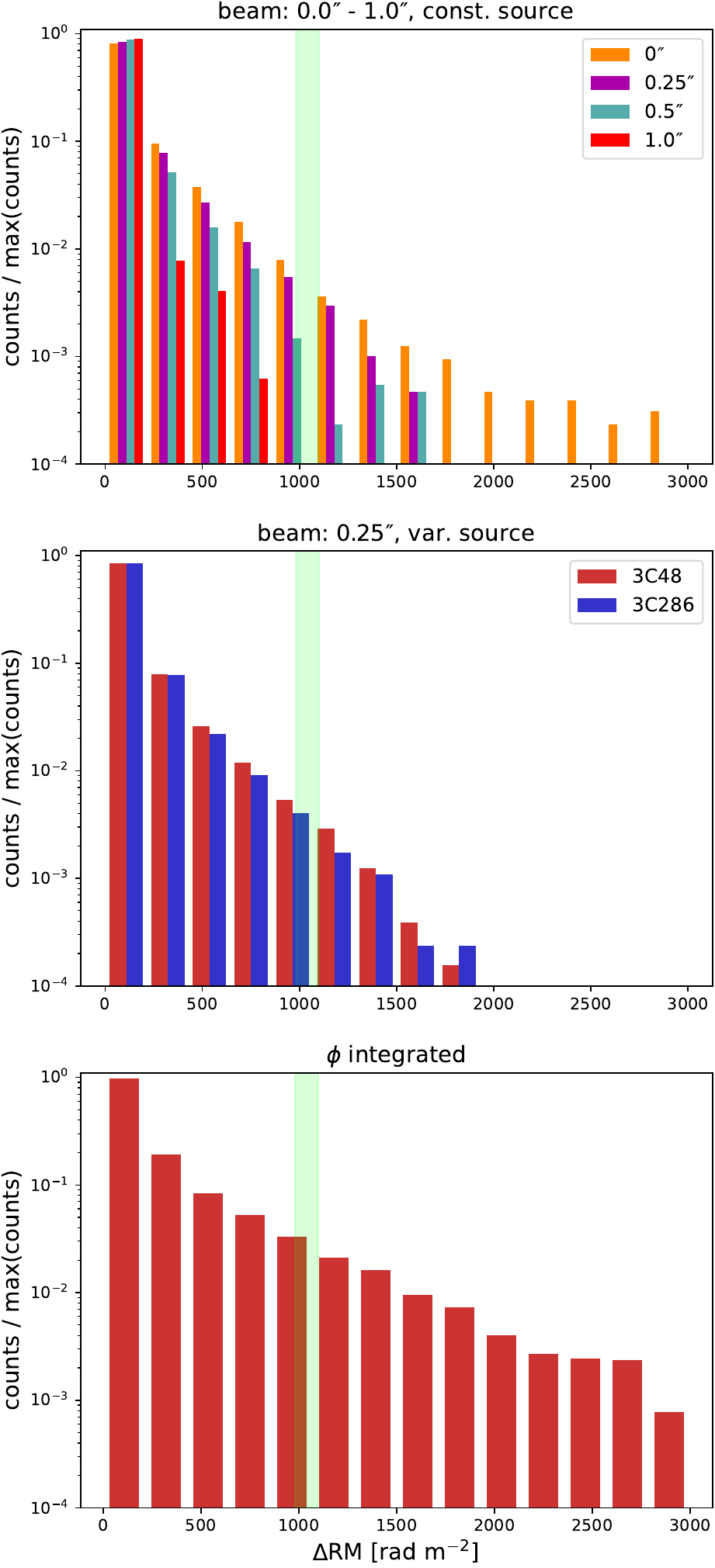}
	\caption{The distribution of the difference in Faraday rotation $\Delta \mathrm{RM}$ between all considered pairs of lensing points LP$_{(-)}$ and LP$_{(+)}$, respectively, for an inclination of ${i=45^\circ}$. Vertical light green shaded areas represent the range of ${ \Delta \mathrm{RM} = 1040\pm 60\ \mathrm{rad\ m}^{-2} }$ reported in \cite{Mao2017NatAs}. Top panel: Color-coded are the results for a beam size of $0\arcsec$ (orange), $0.25\arcsec$ (purple), $0.5\arcsec$ (dark green), and $1.0\arcsec$ (red) for a constant background source. Middle panel: The same as the top right panel, but for a beam size of $0.25\arcsec$, considering the parameters of the sources 3C48 (red) and 3C286 (blue). Bottom panel: The distribution of the Faraday depth $\phi$ of all TNG50 galaxies directly integrated along the LOS of each LP through the cutout. } 
	\label{fig:dRM}
\end{figure}

A key finding of the observations by \cite{Mao2017NatAs} is the reported difference of $\Delta \mathrm{RM}= 1040 \pm 60\ \mathrm{rad\ m^{-2}}$ between two LPs. In this section, we briefly explore the probability of measuring such a value of $\Delta \mathrm{RM}$. The advantage of differential Faraday rotation measurements in lensing systems is that they largely cancel foreground pollutions. Due to their small angular separation, the two images of \textsc{CLASS B1152+199} are expected to experience nearly identical Galactic foreground RMs, eliminating the need for an absolute foreground correction. However, small-scale Galactic fluctuations and differential contributions from the background source or intervening material cannot be excluded. We therefore adopt the rest-frame differential RM reported by \citet{Mao2017NatAs} without propagating additional uncertainties from these possible foreground contributions. 

To do so, we show in Fig.~\ref{fig:dRM} histograms of all the variations in parameters and models as outlined in the previous section. For a frequency independent background source and no beam depolarization, we report values for the difference in RM up to $\Delta \mathrm{RM}=3939\ \mathrm{rad\ m}^{-2}$. However, the occurrence of such high values is rather unlikely, being about three orders of magnitude less frequent than the most likely values, which peak close to $\Delta \mathrm{RM}=0\ \mathrm{rad\ m}^{-2}$. Given the distribution of $\Delta$RM, the probability of finding a $\Delta$RM within the particular range of ${ 1040 \pm 60\ \mathrm{rad\ m^{-2}}}$ is about $0.42\ \%$. With an increasing beam size, the maximal observable $\Delta$RM decreases to $1660\ \mathrm{rad\ m}^{-2}$ for a beam size of $0.5\arcsec$, reaching down to $725\ \mathrm{rad\ m}^{-2}$ for $1.0\arcsec$. The probability of finding $\Delta$RM in the observed interval is $0.05\ \%$ for  a beam with of $0.5\arcsec$, and non-existent for $1.0\arcsec$ and larger. We note that the distribution of $\Delta$RM for variable source-case are rather similar when considering a $0.25\arcsec$ beam. Here, we find probabilities of $0.33\ \%$ and $0.22\ \%$
for 3C48 and 3C286, respectively to recover the actually observed $\Delta$RM range.

Given that our spectral resolution, with a channel width of $34.8\ \mathrm{MHz}$, is larger than the $8\ \mathrm{MHz}$ utilized in \cite{Mao2017NatAs}, this might provide an explanation for the low probabilities of actually observing a difference of ${\Delta \mathrm{RM}= 1040 \pm 60\ \mathrm{rad\ m^{-2}}}$. Naturally, RM synthesis is limited in sensitivity by a maximal value for the Faraday depth $\phi$ of ${ \phi_{\mathrm{max}}=\sqrt{3} / \delta \lambda^2 }$, where ${ \delta \lambda^2 }$ is the difference between two consecutive channels in $\lambda^2$-space (see \citealt{Brentjens2005} and \citealt{Heald2015} for further details). For a channel width of $34.8\ \mathrm{MHz}$, the sensitivity of $\phi$ is limited to ${ \phi_{\mathrm{max}}=2.91 \times 10^2\ \mathrm{rad\ m}^{-2}}$ at $1\ \mathrm{GHz}$ and ${\phi_{\mathrm{max}}=1.41 \times 10^5\ \mathrm{rad\ m}^{-2}}$ at $8\ \mathrm{GHz}$. However, it is not a priori clear how this would limit the measurement of the difference $\Delta$RM over the entire frequency range of $1\ \mathrm{GHz} - 8\ \mathrm{GHz}$, as we see peak values in $\phi$ up to $\approx 6000\ \mathrm{rad\ m}^{-2}$ for individual galaxies and lensing points (compare Fig.~\ref{fig:RM_maps}).

Using the directly integrated Faraday depths, we find that $1.41\ \%$ of all lensing-point pairs reproduce the observed range of $\Delta\mathrm{RM}=1040\pm60\ \mathrm{rad\ m^{-2}}$ \citep{Mao2017NatAs}. Restricting the lensing points to galactocentric distances comparable to the observed $2.6$ and $6.5\ \mathrm{kpc}$ increases this fraction to
$3.16\ \%$.

We conclude that the limit associated with RM synthesis can only partly account for the low probability of finding  a difference of ${\Delta \mathrm{RM} = 1040\pm60\ \mathrm{rad\ m^{-2}}}$. We emphasize that this fraction is conditional on our selected TNG50 galaxies and the adopted lensing-point sampling. Although we restrict the galactocentric distances to values comparable to those observed in \textsc{CLASS B1152+199}, we do not reproduce the full observed lensing geometry.

Moreover, even in the absence of selection effects for \textsc{CLASS B1152+199}, a random sampling of a TNG50 galaxy population could in principle naturally produce a statistical outlier of a few percent, and so the mere existence of such an lensing-system need not necessarily challenge current theoretical models or simulations

In addition to the large magnetic field strengths derived by \citet{Mao2017NatAs} via RM synthesis,  the small-scale variations in the lensing galaxy of the system \textsc{CLASS B1152+199} are also determined using the technique of Stokes QU fitting \citep[see][]{Law2011ApJ}. This technique recovers the small-scale random magnetic field strength typically associated with scales of about ${ 1\ \mathrm{pc} - 10\ \mathrm{pc} }$ \citep{1996ARAA,Beck2015}. However, the TNG50 simulation captures spatial scales only down to $\sim 100\ \mathrm{pc}$ \citep[see][]{Ramesh2024b}, and so smaller-scale magnetic field fluctuations are not resolved. This could also impact the classification of the magnetic field morphology within the disk into dynamo modes of the magnetic field \citep{Beck2007}, and the spatial scale of random field fluctuations in the ISM \citep{Haverkorn2006, Haverkorn2008}. Future high resolution hydrodynamical simulations, including those with explicitly resolved small-scale magnetic field fluctuations, might be able to shed light on these small-scale details \citep{Pfrommer2022,Ramesh2024,Goller2025,Kjellgren2025,Matthew2026,Tress2026}.


\section{Summary and conclusion} \label{sec:summary}

In this paper, we explore constraints on the large-scale magnetic field strength in galaxies at $z \sim 0.5$ available via lensing of the polarized radiation of a background source using multi-frequency Faraday rotation observations. Our investigation takes the system \textsc{CLASS B1152+199} as a reference. For comparison, we select 64 star-forming disk galaxies with stellar masses of ${2\times10^{8}\ M_\odot - 3\times10^{9}\ M_\odot}$ from the TNG50 cosmological magnetohydrodynamical (MHD) simulation of galaxies. We post-process these galaxies with the polarized RT code POLARIS to derive synthetic Faraday observations.

For each lensing galaxy, the POLARIS RT calculations are performed in the Stokes vector formalism, including Faraday rotation based on the electron fractions and magnetic fields within the TNG50 galaxies, as well as the polarized radiation from the background source. The resulting synthetic observations are evaluated by means of RM synthesis to derive the difference $\Delta$RM between pairs of lensing points. To characterize the large-scale magnetic field independently of the total magnetic field strength, we vertically average the signed cylindrical magnetic field components following the approach of \citet{Kovacs2024}. We fit the resulting large-scale field strength with an exponential radial profile and obtain the characteristic field strength $B_{\mathrm{0}}$. Following the observational approach of \citet{Mao2017NatAs}, we then derive an observational estimate $B_{\mathrm{obs}}$ from the synthetic Faraday spectra and compare it with $B_{\mathrm{0}}$. Our main results can be summarized as follows:

\begin{itemize}

\item The signed-component averaging reduces the contribution from small-scale magnetic field fluctuations while retaining the large-scale magnetic field structure without imposing an axis-symmetric (AS) or bisymetrical (BS) field morphology on the TNG50 galaxies. For the entire sample, TNG50 predicts a characteristic large-scale magnetic field strength of $B_{\mathrm{0}} = 10.3\ \mu\mathrm{G}$ and a radial scale length of $r_{\mathrm{MAG}}=8.3\ \mathrm{kpc}$. For the individual galaxies, we find ranges of ${B_{\mathrm{0}} = 1.9\ \mu\mathrm{G} - 15.6\ \mu\mathrm{G}}$ and ${r_{\mathrm{MAG}} = 2.2\ \mathrm{kpc} - 15.5\ \mathrm{kpc}}$, respectively. The resulting ranges are broadly compatible with the magnetic-field strengths and scale lengths considered by \citet{Mao2017NatAs}. However, given the uncertainties in the physical properties of the lensing galaxy, this agreement does not imply that the selected TNG50 galaxies are exact
counterparts of \textsc{CLASS B1152+199}.

\item Assuming AS, BS, or constant geometrical correction factors $C_{\pm}$ results in only minor differences in the inferred magnetic field strength $B_{\mathrm{obs}}$. The AS and BS assumptions yield slightly larger average values of $B_{\mathrm{obs}}$ than the constant case, but the differences remain well within the individual uncertainties. Hence, the assumed large-scale field geometry has only a minor impact on our estimate of $B_{\mathrm{obs}}$ for the considered lensing configurations. Since neither the field morphology of \textsc{CLASS B1152+199} nor that of the individual TNG50 galaxies is known a priori, we therefore adopt constant $C_{\pm}$ for the remaining analysis.

\item Under optimal observational conditions, i.e.\ without beam depolarization and for a well-characterized background source, $B_{\mathrm{obs}}$ correlates well with the characteristic large-scale field strength $B_{\mathrm{0}}$. Depending on galaxy inclination, the inferred values generally recover $B_{\mathrm{0}}$ within a factor of three. For $i=75^\circ$, $83.3\ \%$ of the galaxies fall within ${B_{\mathrm{obs}}/B_{\mathrm{0}}=1-3}$, while for $i=45^\circ$ the probability of finding ${B_{\mathrm{obs}}/B_{\mathrm{0}}}$ within ${1/3-3}$ is $86.3\ \%$. At $i=15^\circ$, $83.5\ \%$ of the galaxies have ratios within ${1/3-1}$. Thus, inclination affects the normalization of the recovered field strength but does not remove the overall correlation between $B_{\mathrm{obs}}$ and $B_{\mathrm{0}}$.

\item Beam depolarization represents the largest observational limitation considered in this work. For a beam size of $0.5\arcsec$, $B_{\mathrm{obs}}$ is typically reduced by approximately a factor of two, while increasing the beam size to $1\arcsec-2\arcsec$ can lead to substantially stronger underestimates. In our extreme $2\arcsec$ beam scenario, the ratio ${B_{\mathrm{obs}}/B_{\mathrm{0}}}$ decreases down to values between approximately $1/10$ and $1/30$. However, such a scenario requires multiple rays to contribute over the entire beam and therefore represents an unlikely limiting case.

\item Despite beam depolarization, the difference in RM recovered through RM synthesis remains clearly correlated with the directly integrated value through the simulation. For a beam size of $0.25\arcsec$ and an inclination of $i=45^\circ$, we find a linear correlation between $\Delta\mathrm{RM}_{\mathrm{obs}}$ and $\Delta\mathrm{RM}_{\mathrm{int}}$ with a slope of $0.496$. Thus, RM synthesis preserves the underlying information on differential Faraday rotation, although the magnitude of $\Delta$RM is systematically reduced under the considered observational conditions.

\item Frequency-dependent polarization of the background source introduces additional broadening and shifts in the Faraday spectra. Nevertheless, when adopting the polarization properties of sources with variable polarization, the average recovered values of $B_{\mathrm{obs}}$ remain consistent within the individual uncertainties. For a beam size of $0.25\arcsec$, between $72.1\ \%$ and $84.8\ \%$ of the inferred field strengths fall within ${B_{\mathrm{obs}}/B_{\mathrm{0}}=1/3-1}$. Hence, frequency-dependent polarization of the background source introduces additional scatter but has a smaller impact on the inferred large-scale field strength than strong beam depolarization.

\item Large differential rotation measures comparable to the value ${\Delta\mathrm{RM}=1040\pm60\ \mathrm{rad\ m}^{-2}}$ observed toward \textsc{CLASS B1152+199} are rare within the selected TNG50 galaxies. Under ideal synthetic observational conditions, only $0.42\ \%$ of all considered lensing-point pairs fall within the observed interval. From the directly integrated Faraday depths, for which limitations associated with RM synthesis are absent, the probability increases to $1.41\ \%$. Restricting the lensing points to galactocentric distances close to $6.5\ \mathrm{kpc}$ and $2.6\ \mathrm{kpc}$, as observed in \textsc{CLASS B1152+199}, increases the probability to $3.16\ \%$. Consequently, limitations of the RM synthesis alone cannot account for the rarity of the observed $\Delta$RM.

\end{itemize}

In conclusion, differential Faraday rotation of gravitationally lensed polarized background sources provides a robust probe of the large-scale magnetic field in intervening galaxies. For the majority of the configurations considered here, the observationally inferred field strength $B_{\mathrm{obs}}$ recovers the characteristic large-scale field strength $B_{\mathrm{0}}$ of the TNG50 galaxies within approximately a factor of three, provided that beam depolarization is moderate. The inferred field strength depends only weakly on the assumed AS or BS field geometry, whereas inclination and, in particular, beam depolarization introduce larger systematic variations.

At the same time, the large differential Faraday rotation observed in \textsc{CLASS B1152+199} lies in the high-$\Delta$RM tail of the TNG50 analog population, even when the observed galactocentric distances of the two lensed images are taken into account. Several factors may contribute to this difference. First, the physical properties of the lensing galaxy remain uncertain, and a more massive or gas-rich system than those represented by most of our analogs could produce larger thermal electron columns and, consequently, larger rotation measures. Second, the finite numerical resolution of TNG50 may affect the distribution of ionized gas and the amplification and structure of magnetic fields, particularly in dense regions of the interstellar medium. Unresolved gas structure and magnetic field amplification may therefore influence the high-$\Delta$RM tail.

Finally, differences in the magneto-ionic structure of the observed galaxy and the simulated analogs, potentially associated with their star formation or feedback histories, could also
contribute to the observed differential Faraday rotation. However, even in the absence of a selection bias for \textsc{CLASS B1152+199}, the small fraction of comparable sightlines found in our TNG50 sample does not by itself imply a significant tension with the simulations. A random sampling of the underlying galaxy population can occasionally select systems from the tails of the distribution, and a probability of a few percent therefore remains statistically plausible. Distinguishing between these possibilities requires better constraints on the lensing galaxy and a systematic investigation of the numerical resolution and physical models used in the simulations. Additional Faraday rotation measurements of gravitational lens systems, together with cosmological MHD simulations resolving smaller-scale magnetic field structure, will therefore be important for determining whether \textsc{CLASS B1152+199} represents an unusually strongly magnetized system or whether such large differential rotation measures are more common among galaxies at $z\sim0.5$.


\section*{Data Availability}

The IllustrisTNG simulations are publicly available and accessible at \url{www.tng-project.org/data} \citep{Nelson2019b}. This work has benefited from the \texttt{scida} analysis library \citep{Byrohl2024}. Data directly related to this publication is available on request from the corresponding authors. 

\begin{acknowledgements}
S.R. and R.S.K. acknowledge funding from the European Research Council in the ERC synergy grant ``{ECOGAL}'' (project ID 855130).  In addition R.S.K. is grateful for funding from DFG and ANR for project ``STARCLUSTERS'' (funding ID KL 1358/22-1). D.N. acknowledges funding from the Deutsche Forschungsgemeinschaft (DFG) through an Emmy Noether Research Group (grant number NE 2441/1-1). The project made use of computing resources provided by {\em The L\"{a}nd} through bwHPC and by DFG through grant  INST 35/1134-1 FUGG and 35/1597-1 FUGG. Data are in part stored at SDS@hd supported by the Ministry of Science, Research and the Arts and by DFG through grant INST 35/1314-1 FUGG and INST 35/1503-1 FUGG. 
\end{acknowledgements}

\bibliographystyle{aa}
\bibliography{./bibtex}

@ARTICLE{Pakmor2018,
       author = {{Pakmor}, R{\"u}diger and {Guillet}, Thomas and {Pfrommer}, Christoph and {G{\'o}mez}, Facundo A. and {Grand}, Robert J.~J. and {Marinacci}, Federico and {Simpson}, Christine M. and {Springel}, Volker},
        title = "{Faraday rotation maps of disc galaxies}",
      journal = {\mnras},
         year = 2018,
        month = dec,
       volume = {481},
       number = {4},
        pages = {4410-4418},
          doi = {10.1093/mnras/sty2601},
archivePrefix = {arXiv},
       eprint = {1807.02113},
 primaryClass = {astro-ph.GA},
       adsurl = {https://ui.adsabs.harvard.edu/abs/2018MNRAS.481.4410P}
}

@ARTICLE{Pakmor2014,
       author = {{Pakmor}, R{\"u}diger and {Marinacci}, Federico and {Springel}, Volker},
        title = "{Magnetic Fields in Cosmological Simulations of Disk Galaxies}",
      journal = {\apjl},
         year = 2014,
        month = mar,
       volume = {783},
       number = {1},
          eid = {L20},
        pages = {L20},
          doi = {10.1088/2041-8205/783/1/L20},
archivePrefix = {arXiv},
       eprint = {1312.2620},
 primaryClass = {astro-ph.GA},
       adsurl = {https://ui.adsabs.harvard.edu/abs/2014ApJ...783L..20P}
}

@ARTICLE{Pakmor2024,
       author = {{Pakmor}, R{\"u}diger and {Bieri}, Rebekka and {van de Voort}, Freeke and {Werhahn}, Maria and {Fattahi}, Azadeh and {Guillet}, Thomas and {Pfrommer}, Christoph and {Springel}, Volker and {Talbot}, Rosie Y.},
        title = "{Magnetic field amplification in cosmological zoom simulations from dwarf galaxies to galaxy groups}",
      journal = {\mnras},
         year = 2024,
        month = feb,
       volume = {528},
       number = {2},
        pages = {2308-2325},
          doi = {10.1093/mnras/stae112},
archivePrefix = {arXiv},
       eprint = {2309.13104},
 primaryClass = {astro-ph.GA},
       adsurl = {https://ui.adsabs.harvard.edu/abs/2024MNRAS.528.2308P}
}

@ARTICLE{Nelson2018,
       author = {{Nelson}, Dylan and {Pillepich}, Annalisa and {Springel}, Volker and {Weinberger}, Rainer and {Hernquist}, Lars and {Pakmor}, R{\"u}diger and {Genel}, Shy and {Torrey}, Paul and {Vogelsberger}, Mark and {Kauffmann}, Guinevere and {Marinacci}, Federico and {Naiman}, Jill},
        title = "{First results from the IllustrisTNG simulations: the galaxy colour bimodality}",
      journal = {\mnras},
         year = 2018,
        month = mar,
       volume = {475},
       number = {1},
        pages = {624-647},
          doi = {10.1093/mnras/stx3040},
archivePrefix = {arXiv},
       eprint = {1707.03395},
 primaryClass = {astro-ph.GA},
       adsurl = {https://ui.adsabs.harvard.edu/abs/2018MNRAS.475..624N}
}

@ARTICLE{Springel2018,
       author = {{Springel}, Volker and {Pakmor}, R{\"u}diger and {Pillepich}, Annalisa and {Weinberger}, Rainer and {Nelson}, Dylan and {Hernquist}, Lars and {Vogelsberger}, Mark and {Genel}, Shy and {Torrey}, Paul and {Marinacci}, Federico and {Naiman}, Jill},
        title = "{First results from the IllustrisTNG simulations: matter and galaxy clustering}",
      journal = {\mnras},
         year = 2018,
        month = mar,
       volume = {475},
       number = {1},
        pages = {676-698},
          doi = {10.1093/mnras/stx3304},
archivePrefix = {arXiv},
       eprint = {1707.03397},
 primaryClass = {astro-ph.GA},
       adsurl = {https://ui.adsabs.harvard.edu/abs/2018MNRAS.475..676S}
}

@ARTICLE{Naiman2018,
       author = {{Naiman}, Jill P. and {Pillepich}, Annalisa and {Springel}, Volker and {Ramirez-Ruiz}, Enrico and {Torrey}, Paul and {Vogelsberger}, Mark and {Pakmor}, R{\"u}diger and {Nelson}, Dylan and {Marinacci}, Federico and {Hernquist}, Lars and {Weinberger}, Rainer and {Genel}, Shy},
        title = "{First results from the IllustrisTNG simulations: a tale of two elements - chemical evolution of magnesium and europium}",
      journal = {\mnras},
         year = 2018,
        month = jun,
       volume = {477},
       number = {1},
        pages = {1206-1224},
          doi = {10.1093/mnras/sty618},
archivePrefix = {arXiv},
       eprint = {1707.03401},
 primaryClass = {astro-ph.GA},
       adsurl = {https://ui.adsabs.harvard.edu/abs/2018MNRAS.477.1206N}
}

@ARTICLE{Rappaz2022,
       author = {{Rappaz}, Yoan and {Schober}, Jennifer and {Girichidis}, Philipp},
        title = "{Rotation measure and synchrotron emission signatures in simulations of magnetized galactic discs}",
      journal = {\mnras},
         year = 2022,
        month = may,
       volume = {512},
       number = {1},
        pages = {1450-1468},
          doi = {10.1093/mnras/stac516},
archivePrefix = {arXiv},
       eprint = {2203.04336},
 primaryClass = {astro-ph.GA},
       adsurl = {https://ui.adsabs.harvard.edu/abs/2022MNRAS.512.1450R}
}

@ARTICLE{Geach2023,
       author = {{Geach}, J.~E. and {Lopez-Rodriguez}, E. and {Doherty}, M.~J. and {Chen}, Jianhang and {Ivison}, R.~J. and {Bendo}, G.~J. and {Dye}, S. and {Coppin}, K.~E.~K.},
        title = "{Polarized thermal emission from dust in a galaxy at redshift 2.6}",
      journal = {\nat},
         year = 2023,
        month = sep,
       volume = {621},
       number = {7979},
        pages = {483-486},
          doi = {10.1038/s41586-023-06346-4},
archivePrefix = {arXiv},
       eprint = {2309.02034},
 primaryClass = {astro-ph.GA},
       adsurl = {https://ui.adsabs.harvard.edu/abs/2023Natur.621..483G}
}

@ARTICLE{Matthew2026,
       author = {{Smith}, Matthew C.},
        title = "{Imladris: a detailed and flexible model for galaxy simulations with individual stars}",
      journal = {\mnras},
         year = 2026,
        month = jul,
       volume = {549},
       number = {3},
          eid = {stag959},
        pages = {stag959},
          doi = {10.1093/mnras/stag959},
archivePrefix = {arXiv},
       eprint = {2604.00100},
 primaryClass = {astro-ph.GA},
       adsurl = {https://ui.adsabs.harvard.edu/abs/2026MNRAS.549ag959S}
}

@ARTICLE{Pfrommer2022,
       author = {{Pfrommer}, Christoph and {Werhahn}, Maria and {Pakmor}, R{\"u}diger and {Girichidis}, Philipp and {Simpson}, Christine M.},
        title = "{Simulating radio synchrotron emission in star-forming galaxies: small-scale magnetic dynamo and the origin of the far-infrared-radio correlation}",
      journal = {\mnras},
         year = 2022,
        month = sep,
       volume = {515},
       number = {3},
        pages = {4229-4264},
          doi = {10.1093/mnras/stac1808},
archivePrefix = {arXiv},
       eprint = {2105.12132},
 primaryClass = {astro-ph.GA},
       adsurl = {https://ui.adsabs.harvard.edu/abs/2022MNRAS.515.4229P}
}

@ARTICLE{Tress2026,
       author = {{Tress}, R.~G. and {Brucy}, N. and {Girichidis}, P. and {Glover}, S.~C.~O. and {G{\"o}ller}, J. and {Hirschmann}, M. and {Klessen}, R.~S. and {Peter}, T. and {Petersson}, J. and {Sormani}, M.~C. and {Armillotta}, L. and {Battersby}, C.~D. and {Donati}, M. and {Feng}, Z.-X. and {Henshaw}, J.~D. and {Lipman}, D.~R. and {Longmore}, S.~N. and {Nogueras-Lara}, F. and {Pelkonen}, V.-M. and {Peschken}, N. and {Petkova}, M.~A. and {Plat}, A. and {Reissl}, S. and {Smith}, R. and {Soler}, J.~D.},
        title = "{Rhea-RT: Dynamical impact of Central Molecular Zone conditions on the properties of the interstellar medium and stellar feedback coupling}",
      journal = {\aap},
         year = 2026,
        month = jul,
       volume = {711},
          eid = {A103},
        pages = {A103},
          doi = {10.1051/0004-6361/202558503},
archivePrefix = {arXiv},
       eprint = {2512.09981},
 primaryClass = {astro-ph.GA},
       adsurl = {https://ui.adsabs.harvard.edu/abs/2026A&A...711A.103T}
}

@ARTICLE{Goller2025,
       author = {{G{\"o}ller}, Junia and {Girichidis}, Philipp and {Brucy}, No{\'e} and {Hunter}, Glen and {Kjellgren}, Karin and {Tress}, Robin and {Klessen}, Ralf S. and {Glover}, Simon C.~O. and {Hennebelle}, Patrick and {Molinari}, Sergio and {Smith}, Rowan and {Soler}, Juan D. and {Sormani}, Mattia C. and {Testi}, Leonardo},
        title = "{Introducing the Rhea simulations of Milky Way-like galaxies: I. Effect of gravitational potential on morphology and star formation}",
      journal = {\aap},
         year = 2025,
        month = dec,
       volume = {704},
          eid = {A331},
        pages = {A331},
          doi = {10.1051/0004-6361/202452223},
archivePrefix = {arXiv},
       eprint = {2502.02646},
 primaryClass = {astro-ph.GA},
       adsurl = {https://ui.adsabs.harvard.edu/abs/2025A&A...704A.331G}
}

@ARTICLE{Kjellgren2025,
       author = {{Kjellgren}, Karin and {Girichidis}, Philipp and {G{\"o}ller}, Junia and {Brucy}, No{\'e} and {Klessen}, Ralf S. and {Tress}, Robin G. and {Soler}, Juan D. and {Pfrommer}, Christoph and {Werhahn}, Maria and {Glover}, Simon C.~O. and {Smith}, Rowan and {Testi}, Leonardo and {Molinari}, Sergio},
        title = "{The dynamical impact of cosmic rays in the Rhea magnetohydrodynamic simulations}",
      journal = {\aap},
         year = 2025,
        month = aug,
       volume = {700},
          eid = {A124},
        pages = {A124},
          doi = {10.1051/0004-6361/202553754},
archivePrefix = {arXiv},
       eprint = {2502.02635},
 primaryClass = {astro-ph.GA},
       adsurl = {https://ui.adsabs.harvard.edu/abs/2025A&A...700A.124K}
}

@ARTICLE{Pillepich2024,
       author = {{Pillepich}, Annalisa and {Sotillo-Ramos}, Diego and {Ramesh}, Rahul and {Nelson}, Dylan and {Engler}, Christoph and {Rodriguez-Gomez}, Vicente and {Fournier}, Martin and {Donnari}, Martina and {Springel}, Volker and {Hernquist}, Lars},
        title = "{Milky Way and Andromeda analogues from the TNG50 simulation}",
      journal = {\mnras},
         year = 2024,
        month = dec,
       volume = {535},
       number = {2},
        pages = {1721-1762},
          doi = {10.1093/mnras/stae2165},
archivePrefix = {arXiv},
       eprint = {2303.16217},
 primaryClass = {astro-ph.GA},
       adsurl = {https://ui.adsabs.harvard.edu/abs/2024MNRAS.535.1721P}
}

@ARTICLE{Reissl2026,
    author = {Reissl, Stefan and others},
    title  = {Title of the work},
    year   = {2026},
    note   = {in preparation}
}

@ARTICLE{Mao2017NatAs,
       author = {{Mao}, S.~A. and {Carilli}, C. and {Gaensler}, B.~M. and {Wucknitz}, O. and {Keeton}, C. and {Basu}, A. and {Beck}, R. and {Kronberg}, P.~P. and {Zweibel}, E.},
        title = "{Detection of microgauss coherent magnetic fields in a galaxy five billion years ago}",
      journal = {Nature Astronomy},
         year = 2017,
        month = aug,
       volume = {1},
        pages = {621-626},
          doi = {10.1038/s41550-017-0218-x},
archivePrefix = {arXiv},
       eprint = {1708.07844},
 primaryClass = {astro-ph.GA},
       adsurl = {https://ui.adsabs.harvard.edu/abs/2017NatAs...1..621M}
}

@ARTICLE{Reissl2016,
       author = {{Reissl}, S. and {Wolf}, S. and {Brauer}, R.},
        title = "{Radiative transfer with POLARIS. I. Analysis of magnetic fields through synthetic dust continuum polarization measurements}",
      journal = {\aap},
         year = 2016,
        month = sep,
       volume = {593},
          eid = {A87},
        pages = {A87},
          doi = {10.1051/0004-6361/201424930},
archivePrefix = {arXiv},
       eprint = {1604.05305},
 primaryClass = {astro-ph.IM},
       adsurl = {https://ui.adsabs.harvard.edu/abs/2016A&A...593A..87R}
}

@MISC{Purcell2020,
       author = {{Purcell}, C.~R. and {Van Eck}, C.~L. and {West}, J. and {Sun}, X.~H. and {Gaensler}, B.~M.},
        title = "{RM-Tools: Rotation measure (RM) synthesis and Stokes QU-fitting}",
 howpublished = {Astrophysics Source Code Library, record ascl:2005.003},
         year = 2020,
        month = may,
          eid = {ascl:2005.003},
        pages = {ascl:2005.003},
archivePrefix = {ascl},
       eprint = {2005.003},
       adsurl = {https://ui.adsabs.harvard.edu/abs/2020ascl.soft05003P}
}

@ARTICLE{Beck2015,
       author = {{Beck}, Rainer},
        title = "{Magnetic fields in spiral galaxies}",
      journal = {\aapr},
         year = 2015,
        month = dec,
       volume = {24},
          eid = {4},
        pages = {4},
          doi = {10.1007/s00159-015-0084-4},
archivePrefix = {arXiv},
       eprint = {1509.04522},
 primaryClass = {astro-ph.GA},
       adsurl = {https://ui.adsabs.harvard.edu/abs/2015A&ARv..24....4B}
}

@ARTICLE{Perley2013a,
       author = {{Perley}, R.~A. and {Butler}, B.~J.},
        title = "{An Accurate Flux Density Scale from 1 to 50 GHz}",
      journal = {\apjs},
         year = 2013,
        month = feb,
       volume = {204},
       number = {2},
          eid = {19},
        pages = {19},
          doi = {10.1088/0067-0049/204/2/19},
archivePrefix = {arXiv},
       eprint = {1211.1300},
 primaryClass = {astro-ph.IM},
       adsurl = {https://ui.adsabs.harvard.edu/abs/2013ApJS..204...19P}
}

@ARTICLE{Perley2013b,
       author = {{Perley}, R.~A. and {Butler}, B.~J.},
        title = "{Integrated Polarization Properties of 3C48, 3C138, 3C147, and 3C286}",
      journal = {\apjs},
         year = 2013,
        month = jun,
       volume = {206},
       number = {2},
          eid = {16},
        pages = {16},
          doi = {10.1088/0067-0049/206/2/16},
archivePrefix = {arXiv},
       eprint = {1302.6662},
 primaryClass = {astro-ph.IM},
       adsurl = {https://ui.adsabs.harvard.edu/abs/2013ApJS..206...16P}
}

@book{Bohren1998,
  author    = {{Bohren}, Craig F. and {Huffman}, Donald R.},
  title     = {{Absorption and Scattering of Light by Small Particles}},
  year      = 1998,
  publisher = {Wiley-VCH},
  adsurl    = {https://ui.adsabs.harvard.edu/abs/1998asls.book.....B}
}

@BOOK{Rybicki1986,
       author = {{Rybicki}, George B. and {Lightman}, Alan P.},
        title = "{Radiative Processes in Astrophysics}",
         year = 1986,
       adsurl = {https://ui.adsabs.harvard.edu/abs/1986rpa..book.....R}
}

@ARTICLE{Sokoloff1998,
       author = {{Sokoloff}, D.~D. and {Bykov}, A.~A. and {Shukurov}, A. and {Berkhuijsen}, E.~M. and {Beck}, R. and {Poezd}, A.~D.},
        title = "{Depolarization and Faraday effects in galaxies}",
      journal = {\mnras},
         year = 1998,
        month = aug,
       volume = {299},
       number = {1},
        pages = {189-206},
          doi = {10.1046/j.1365-8711.1998.01782.x},
       adsurl = {https://ui.adsabs.harvard.edu/abs/1998MNRAS.299..189S}
}

@ARTICLE{Beck2009,
       author = {{Beck}, R.},
        title = "{Galactic and extragalactic magnetic fields - a concise review}",
      journal = {Astrophysics and Space Sciences Transactions},
         year = 2009,
        month = oct,
       volume = {5},
       number = {1},
        pages = {43-47},
          doi = {10.5194/astra-5-43-2009},
       adsurl = {https://ui.adsabs.harvard.edu/abs/2009ASTRA...5...43B}
}

@ARTICLE{Heesen2023,
       author = {{Heesen}, V. and {Klocke}, T.-L. and {Br{\"u}ggen}, M. and {Tabatabaei}, F.~S. and {Basu}, A. and {Beck}, R. and {Drabent}, A. and {Nikiel-Wroczy{\'n}ski}, B. and {Paladino}, R. and {Schulz}, S. and {Stein}, M.},
        title = "{Nearby galaxies in the LOFAR Two-metre Sky Survey. II. The magnetic field-gas relation}",
      journal = {\aap},
         year = 2023,
        month = jan,
       volume = {669},
          eid = {A8},
        pages = {A8},
          doi = {10.1051/0004-6361/202243328},
archivePrefix = {arXiv},
       eprint = {2208.11068},
 primaryClass = {astro-ph.GA},
       adsurl = {https://ui.adsabs.harvard.edu/abs/2023A&A...669A...8H}
}

@ARTICLE{Shcherbakov2008,
   author = {{Shcherbakov}, R.~V.},
    title = "{Propagation Effects in Magnetized Transrelativistic Plasmas}",
  journal = {\apj},
archivePrefix = "arXiv",
   eprint = {0809.0012},
     year = 2008,
    month = nov,
   volume = 688,
    pages = {695-700},
      doi = {10.1086/592326},
   adsurl = {http://esoads.eso.org/abs/2008ApJ...688..695S}
}

@ARTICLE{Pandya2016,
   author = {{Pandya}, A. and {Zhang}, Z. and {Chandra}, M. and {Gammie}, C.~F.
	},
    title = "{Polarized Synchrotron Emissivities and Absorptivities for Relativistic Thermal, Power-law, and Kappa Distribution Functions}",
  journal = {\apj},
archivePrefix = "arXiv",
   eprint = {1602.08749},
 primaryClass = "astro-ph.HE",
     year = 2016,
    month = may,
   volume = 822,
      eid = {34},
    pages = {34},
      doi = {10.3847/0004-637X/822/1/34},
   adsurl = {http://esoads.eso.org/abs/2016ApJ...822...34P}
}

@ARTICLE{Dexter2016,
   author = {{Dexter}, J.},
    title = "{A public code for general relativistic, polarised radiative transfer around spinning black holes}",
  journal = {\mnras},
archivePrefix = "arXiv",
   eprint = {1602.03184},
 primaryClass = "astro-ph.HE",
     year = 2016,
    month = oct,
   volume = 462,
    pages = {115-136},
      doi = {10.1093/mnras/stw1526},
   adsurl = {http://esoads.eso.org/abs/2016MNRAS.462..115D}
}

@ARTICLE{Reissl2019,
       author = {{Reissl}, Stefan and {Brauer}, Robert and {Klessen}, Ralf S. and {Pellegrini}, Eric W.},
        title = "{Radiative Transfer with POLARIS. II. Modeling of Synthetic Galactic Synchrotron Observations}",
      journal = {\apj},
         year = 2019,
        month = nov,
       volume = {885},
       number = {1},
          eid = {15},
        pages = {15},
          doi = {10.3847/1538-4357/ab3664},
archivePrefix = {arXiv},
       eprint = {1907.12617},
 primaryClass = {astro-ph.GA},
       adsurl = {https://ui.adsabs.harvard.edu/abs/2019ApJ...885...15R}
}

@BOOK{Ruzmaikin1988,
       author = {{Ruzmaikin}, Aleksandr A. and {Sokolov}, Dmitrii D. and {Shukurov}, Anvar M.},
        title = "{Magnetic Fields of Galaxies}",
         year = 1988,
       volume = {133},
          doi = {10.1007/978-94-009-2835-0},
       adsurl = {https://ui.adsabs.harvard.edu/abs/1988ASSL..133.....R}
}

@ARTICLE{Myers1999,
       author = {{Myers}, S.~T. and {Rusin}, D. and {Fassnacht}, C.~D. and {Blandford}, R.~D. and {Pearson}, T.~J. and {Readhead}, A.~C.~S. and {Jackson}, N. and {Browne}, I.~W.~A. and {Marlow}, D.~R. and {Wilkinson}, P.~N. and {Koopmans}, L.~V.~E. and {de Bruyn}, A.~G.},
        title = "{CLASS B1152+199 and B1359+154: Two New Gravitational Lens Systems Discovered in the Cosmic Lens All-Sky Survey}",
      journal = {\aj},
         year = 1999,
        month = jun,
       volume = {117},
       number = {6},
        pages = {2565-2572},
          doi = {10.1086/300875},
archivePrefix = {arXiv},
       eprint = {astro-ph/9905043},
 primaryClass = {astro-ph},
       adsurl = {https://ui.adsabs.harvard.edu/abs/1999AJ....117.2565M}
}

@ARTICLE{Rusin2002,
       author = {{Rusin}, D. and {Norbury}, M. and {Biggs}, A.~D. and {Marlow}, D.~R. and {Jackson}, N.~J. and {Browne}, I.~W.~A. and {Wilkinson}, P.~N. and {Myers}, S.~T.},
        title = "{High-resolution observations and mass modelling of the CLASS gravitational lens B1152+199}",
      journal = {\mnras},
         year = 2002,
        month = feb,
       volume = {330},
       number = {1},
        pages = {205-211},
          doi = {10.1046/j.1365-8711.2002.05043.x},
archivePrefix = {arXiv},
       eprint = {astro-ph/0110099},
 primaryClass = {astro-ph},
       adsurl = {https://ui.adsabs.harvard.edu/abs/2002MNRAS.330..205R}
}

@ARTICLE{Dai2009,
       author = {{Dai}, Xinyu and {Kochanek}, Christopher S.},
        title = "{Differential X-Ray Absorption and Dust-to-Gas Ratios of the Lens Galaxies SBS 0909+523, FBQS 0951+2635, and B 1152+199}",
      journal = {\apj},
         year = 2009,
        month = feb,
       volume = {692},
       number = {1},
        pages = {677-683},
          doi = {10.1088/0004-637X/692/1/677},
archivePrefix = {arXiv},
       eprint = {0803.1679},
 primaryClass = {astro-ph},
       adsurl = {https://ui.adsabs.harvard.edu/abs/2009ApJ...692..677D}
}

@ARTICLE{Toft2000,
       author = {{Toft}, S. and {Hjorth}, J. and {Burud}, I.},
        title = "{The extinction curve of the lensing galaxy of B1152+199 at z=0.44}",
      journal = {\aap},
         year = 2000,
        month = may,
       volume = {357},
        pages = {115-119},
          doi = {10.48550/arXiv.astro-ph/0003098},
archivePrefix = {arXiv},
       eprint = {astro-ph/0003098},
 primaryClass = {astro-ph},
       adsurl = {https://ui.adsabs.harvard.edu/abs/2000A&A...357..115T}
}

@ARTICLE{Kulsrud2008,
       author = {{Kulsrud}, Russell M. and {Zweibel}, Ellen G.},
        title = "{On the origin of cosmic magnetic fields}",
      journal = {Reports on Progress in Physics},
         year = 2008,
        month = apr,
       volume = {71},
       number = {4},
          eid = {046901},
        pages = {046901},
          doi = {10.1088/0034-4885/71/4/046901},
archivePrefix = {arXiv},
       eprint = {0707.2783},
 primaryClass = {astro-ph},
       adsurl = {https://ui.adsabs.harvard.edu/abs/2008RPPh...71d6901K}
}

@ARTICLE{Pillepich2018a,
       author = {{Pillepich}, Annalisa and {Springel}, Volker and {Nelson}, Dylan and {Genel}, Shy and {Naiman}, Jill and {Pakmor}, R{\"u}diger and {Hernquist}, Lars and {Torrey}, Paul and {Vogelsberger}, Mark and {Weinberger}, Rainer and {Marinacci}, Federico},
        title = "{Simulating galaxy formation with the IllustrisTNG model}",
      journal = {\mnras},
         year = 2018,
        month = jan,
       volume = {473},
       number = {3},
        pages = {4077-4106},
          doi = {10.1093/mnras/stx2656},
archivePrefix = {arXiv},
       eprint = {1703.02970},
 primaryClass = {astro-ph.GA},
       adsurl = {https://ui.adsabs.harvard.edu/abs/2018MNRAS.473.4077P}
}

@ARTICLE{Weinberger2017,
       author = {{Weinberger}, Rainer and {Springel}, Volker and {Hernquist}, Lars and {Pillepich}, Annalisa and {Marinacci}, Federico and {Pakmor}, R{\"u}diger and {Nelson}, Dylan and {Genel}, Shy and {Vogelsberger}, Mark and {Naiman}, Jill and {Torrey}, Paul},
        title = "{Simulating galaxy formation with black hole driven thermal and kinetic feedback}",
      journal = {\mnras},
         year = 2017,
        month = mar,
       volume = {465},
       number = {3},
        pages = {3291-3308},
          doi = {10.1093/mnras/stw2944},
archivePrefix = {arXiv},
       eprint = {1607.03486},
 primaryClass = {astro-ph.GA},
       adsurl = {https://ui.adsabs.harvard.edu/abs/2017MNRAS.465.3291W}
}

@ARTICLE{Springel2003,
       author = {{Springel}, Volker and {Hernquist}, Lars},
        title = "{Cosmological smoothed particle hydrodynamics simulations: a hybrid multiphase model for star formation}",
      journal = {\mnras},
         year = 2003,
        month = feb,
       volume = {339},
       number = {2},
        pages = {289-311},
          doi = {10.1046/j.1365-8711.2003.06206.x},
archivePrefix = {arXiv},
       eprint = {astro-ph/0206393},
 primaryClass = {astro-ph},
       adsurl = {https://ui.adsabs.harvard.edu/abs/2003MNRAS.339..289S}
}

@BOOK{Peebles1993,
       author = {{Peebles}, P.~J.~E.},
        title = "{Principles of Physical Cosmology}",
         year = 1993,
          doi = {10.1515/9780691206721},
       adsurl = {https://ui.adsabs.harvard.edu/abs/1993ppc..book.....P}
}

@ARTICLE{Marinacci2018,
       author = {{Marinacci}, Federico and {Vogelsberger}, Mark and {Pakmor}, R{\"u}diger and {Torrey}, Paul and {Springel}, Volker and {Hernquist}, Lars and {Nelson}, Dylan and {Weinberger}, Rainer and {Pillepich}, Annalisa and {Naiman}, Jill and {Genel}, Shy},
        title = "{First results from the IllustrisTNG simulations: radio haloes and magnetic fields}",
      journal = {\mnras},
         year = 2018,
        month = nov,
       volume = {480},
       number = {4},
        pages = {5113-5139},
          doi = {10.1093/mnras/sty2206},
archivePrefix = {arXiv},
       eprint = {1707.03396},
 primaryClass = {astro-ph.CO},
       adsurl = {https://ui.adsabs.harvard.edu/abs/2018MNRAS.480.5113M}
}

@ARTICLE{Gutierrez2010,
       author = {{Guti{\'e}rrez}, Leonel and {Beckman}, John E.},
        title = "{The Galaxy-wide Distributions of Mean Electron Density in the H II Regions of M51 and NGC 4449}",
      journal = {\apjl},
         year = 2010,
        month = feb,
       volume = {710},
       number = {1},
        pages = {L44-L48},
          doi = {10.1088/2041-8205/710/1/L44},
archivePrefix = {arXiv},
       eprint = {1001.0276},
 primaryClass = {astro-ph.CO},
       adsurl = {https://ui.adsabs.harvard.edu/abs/2010ApJ...710L..44G}
}

@INPROCEEDINGS{Heald2009,
       author = {{Heald}, George},
        title = "{The Faraday rotation measure synthesis technique}",
    booktitle = {Cosmic Magnetic Fields: From Planets, to Stars and Galaxies},
         year = 2009,
       editor = {{Strassmeier}, Klaus G. and {Kosovichev}, Alexander G. and {Beckman}, John E.},
       series = {IAU Symposium},
       volume = {259},
        month = apr,
        pages = {591-602},
          doi = {10.1017/S1743921309031421},
       adsurl = {https://ui.adsabs.harvard.edu/abs/2009IAUS..259..591H}
}

@ARTICLE{Hogbom1974,
       author = {{H{\"o}gbom}, J.~A.},
        title = "{Aperture Synthesis with a Non-Regular Distribution of Interferometer Baselines}",
      journal = {\aaps},
         year = 1974,
        month = jun,
       volume = {15},
        pages = {417},
       adsurl = {https://ui.adsabs.harvard.edu/abs/1974A&AS...15..417H}
}

@ARTICLE{Springel2001,
       author = {{Springel}, Volker and {White}, Simon D.~M. and {Tormen}, Giuseppe and {Kauffmann}, Guinevere},
        title = "{Populating a cluster of galaxies - I. Results at z=0}",
      journal = {\mnras},
         year = 2001,
        month = dec,
       volume = {328},
       number = {3},
        pages = {726-750},
          doi = {10.1046/j.1365-8711.2001.04912.x},
archivePrefix = {arXiv},
       eprint = {astro-ph/0012055},
 primaryClass = {astro-ph},
       adsurl = {https://ui.adsabs.harvard.edu/abs/2001MNRAS.328..726S}
}

@ARTICLE{Pakmor2017,
       author = {{Pakmor}, R{\"u}diger and {G{\'o}mez}, Facundo A. and {Grand}, Robert J.~J. and {Marinacci}, Federico and {Simpson}, Christine M. and {Springel}, Volker and {Campbell}, David J.~R. and {Frenk}, Carlos S. and {Guillet}, Thomas and {Pfrommer}, Christoph and {White}, Simon D.~M.},
        title = "{Magnetic field formation in the Milky Way like disc galaxies of the Auriga project}",
      journal = {\mnras},
         year = 2017,
        month = aug,
       volume = {469},
       number = {3},
        pages = {3185-3199},
          doi = {10.1093/mnras/stx1074},
archivePrefix = {arXiv},
       eprint = {1701.07028},
 primaryClass = {astro-ph.GA},
       adsurl = {https://ui.adsabs.harvard.edu/abs/2017MNRAS.469.3185P}
}

@ARTICLE{Nelson2019b,
       author = {{Nelson}, Dylan and {Springel}, Volker and {Pillepich}, Annalisa and {Rodriguez-Gomez}, Vicente and {Torrey}, Paul and {Genel}, Shy and {Vogelsberger}, Mark and {Pakmor}, Ruediger and {Marinacci}, Federico and {Weinberger}, Rainer and {Kelley}, Luke and {Lovell}, Mark and {Diemer}, Benedikt and {Hernquist}, Lars},
        title = "{The IllustrisTNG simulations: public data release}",
      journal = {Computational Astrophysics and Cosmology},
         year = 2019,
        month = may,
       volume = {6},
       number = {1},
          eid = {2},
        pages = {2},
          doi = {10.1186/s40668-019-0028-x},
archivePrefix = {arXiv},
       eprint = {1812.05609},
 primaryClass = {astro-ph.GA},
       adsurl = {https://ui.adsabs.harvard.edu/abs/2019ComAC...6....2N}
}

@ARTICLE{Nelson2019,
       author = {{Nelson}, Dylan and {Pillepich}, Annalisa and {Springel}, Volker and {Pakmor}, R{\"u}diger and {Weinberger}, Rainer and {Genel}, Shy and {Torrey}, Paul and {Vogelsberger}, Mark and {Marinacci}, Federico and {Hernquist}, Lars},
        title = "{First results from the TNG50 simulation: galactic outflows driven by supernovae and black hole feedback}",
      journal = {\mnras},
         year = 2019,
        month = dec,
       volume = {490},
       number = {3},
        pages = {3234-3261},
          doi = {10.1093/mnras/stz2306},
archivePrefix = {arXiv},
       eprint = {1902.05554},
 primaryClass = {astro-ph.GA},
       adsurl = {https://ui.adsabs.harvard.edu/abs/2019MNRAS.490.3234N}
}

@ARTICLE{Ramesh2023,
       author = {{Ramesh}, Rahul and {Nelson}, Dylan and {Heesen}, Volker and {Br{\"u}ggen}, Marcus},
        title = "{Azimuthal anisotropy of magnetic fields in the circumgalactic medium driven by galactic feedback processes}",
      journal = {\mnras},
         year = 2023,
        month = dec,
       volume = {526},
       number = {4},
        pages = {5483-5493},
          doi = {10.1093/mnras/stad3104},
archivePrefix = {arXiv},
       eprint = {2305.11214},
 primaryClass = {astro-ph.GA},
       adsurl = {https://ui.adsabs.harvard.edu/abs/2023MNRAS.526.5483R}
}

@ARTICLE{Nelson2020,
       author = {{Nelson}, Dylan and {Sharma}, Prateek and {Pillepich}, Annalisa and {Springel}, Volker and {Pakmor}, R{\"u}diger and {Weinberger}, Rainer and {Vogelsberger}, Mark and {Marinacci}, Federico and {Hernquist}, Lars},
        title = "{Resolving small-scale cold circumgalactic gas in TNG50}",
      journal = {\mnras},
         year = 2020,
        month = oct,
       volume = {498},
       number = {2},
        pages = {2391-2414},
          doi = {10.1093/mnras/staa2419},
archivePrefix = {arXiv},
       eprint = {2005.09654},
 primaryClass = {astro-ph.GA},
       adsurl = {https://ui.adsabs.harvard.edu/abs/2020MNRAS.498.2391N}
}

@ARTICLE{Ramesh2024,
       author = {{Ramesh}, Rahul and {Nelson}, Dylan and {Fielding}, Drummond and {Br{\"u}ggen}, Marcus},
        title = "{Zooming in on the circumgalactic medium with GIBLE. The topology and draping of magnetic fields around cold clouds}",
      journal = {\aap},
         year = 2024,
        month = apr,
       volume = {684},
          eid = {L16},
        pages = {L16},
          doi = {10.1051/0004-6361/202348786},
archivePrefix = {arXiv},
       eprint = {2404.01370},
 primaryClass = {astro-ph.GA},
       adsurl = {https://ui.adsabs.harvard.edu/abs/2024A&A...684L..16R}
}

@ARTICLE{Byrohl2024,
       author = {{Byrohl}, Chris and {Nelson}, Dylan},
        title = "{scida: scalable analysis for scientific big data}",
      journal = {The Journal of Open Source Software},
         year = 2024,
        month = feb,
       volume = {9},
       number = {94},
          eid = {6064},
        pages = {6064},
          doi = {10.21105/joss.06064},
archivePrefix = {arXiv},
       eprint = {2402.17818},
 primaryClass = {astro-ph.IM},
       adsurl = {https://ui.adsabs.harvard.edu/abs/2024JOSS....9.6064B}
}

@ARTICLE{Ramesh2024b,
       author = {{Ramesh}, Rahul and {Nelson}, Dylan},
        title = "{Zooming in on the circumgalactic medium with GIBLE: Resolving small-scale gas structure in cosmological simulations}",
      journal = {\mnras},
         year = 2024,
        month = feb,
       volume = {528},
       number = {2},
        pages = {3320-3339},
          doi = {10.1093/mnras/stae237},
archivePrefix = {arXiv},
       eprint = {2307.11143},
 primaryClass = {astro-ph.GA},
       adsurl = {https://ui.adsabs.harvard.edu/abs/2024MNRAS.528.3320R}
}

@ARTICLE{Aramburo2021,
       author = {{Ar{\'a}mburo-Garc{\'\i}a}, Andr{\'e}s and {Bondarenko}, Kyrylo and {Boyarsky}, Alexey and {Nelson}, Dylan and {Pillepich}, Annalisa and {Sokolenko}, Anastasia},
        title = "{Magnetization of the intergalactic medium in the IllustrisTNG simulations: the importance of extended, outflow-driven bubbles}",
      journal = {\mnras},
         year = 2021,
        month = aug,
       volume = {505},
       number = {4},
        pages = {5038-5057},
          doi = {10.1093/mnras/stab1632},
archivePrefix = {arXiv},
       eprint = {2011.11581},
 primaryClass = {astro-ph.CO},
       adsurl = {https://ui.adsabs.harvard.edu/abs/2021MNRAS.505.5038A}
}

@ARTICLE{Pillepich2019,
       author = {{Pillepich}, Annalisa and {Nelson}, Dylan and {Springel}, Volker and {Pakmor}, R{\"u}diger and {Torrey}, Paul and {Weinberger}, Rainer and {Vogelsberger}, Mark and {Marinacci}, Federico and {Genel}, Shy and {van der Wel}, Arjen and {Hernquist}, Lars},
        title = "{First results from the TNG50 simulation: the evolution of stellar and gaseous discs across cosmic time}",
      journal = {\mnras},
         year = 2019,
        month = dec,
       volume = {490},
       number = {3},
        pages = {3196-3233},
          doi = {10.1093/mnras/stz2338},
archivePrefix = {arXiv},
       eprint = {1902.05553},
 primaryClass = {astro-ph.GA},
       adsurl = {https://ui.adsabs.harvard.edu/abs/2019MNRAS.490.3196P}
}

@ARTICLE{Springel2010,
       author = {{Springel}, Volker},
        title = "{E pur si muove: Galilean-invariant cosmological hydrodynamical simulations on a moving mesh}",
      journal = {\mnras},
         year = 2010,
        month = jan,
       volume = {401},
       number = {2},
        pages = {791-851},
          doi = {10.1111/j.1365-2966.2009.15715.x},
archivePrefix = {arXiv},
       eprint = {0901.4107},
 primaryClass = {astro-ph.CO},
       adsurl = {https://ui.adsabs.harvard.edu/abs/2010MNRAS.401..791S}
}

@ARTICLE{Pillepich2018,
       author = {{Pillepich}, Annalisa and {Nelson}, Dylan and {Hernquist}, Lars and {Springel}, Volker and {Pakmor}, R{\"u}diger and {Torrey}, Paul and {Weinberger}, Rainer and {Genel}, Shy and {Naiman}, Jill P. and {Marinacci}, Federico and {Vogelsberger}, Mark},
        title = "{First results from the IllustrisTNG simulations: the stellar mass content of groups and clusters of galaxies}",
      journal = {\mnras},
         year = 2018,
        month = mar,
       volume = {475},
       number = {1},
        pages = {648-675},
          doi = {10.1093/mnras/stx3112},
archivePrefix = {arXiv},
       eprint = {1707.03406},
 primaryClass = {astro-ph.GA},
       adsurl = {https://ui.adsabs.harvard.edu/abs/2018MNRAS.475..648P}
}

@ARTICLE{Furlanetto2006,
       author = {{Furlanetto}, Steven R. and {McQuinn}, Matthew and {Hernquist}, Lars},
        title = "{Characteristic scales during reionization}",
      journal = {\mnras},
         year = 2006,
        month = jan,
       volume = {365},
       number = {1},
        pages = {115-126},
          doi = {10.1111/j.1365-2966.2005.09687.x},
archivePrefix = {arXiv},
       eprint = {astro-ph/0507524},
 primaryClass = {astro-ph},
       adsurl = {https://ui.adsabs.harvard.edu/abs/2006MNRAS.365..115F}
}

@ARTICLE{1996ARAA,
       author = {{Beck}, Rainer and {Brandenburg}, Axel and {Moss}, David and {Shukurov}, Anvar and {Sokoloff}, Dmitry},
        title = "{Galactic Magnetism: Recent Developments and Perspectives}",
      journal = {\araa},
         year = 1996,
        month = jan,
       volume = {34},
        pages = {155-206},
          doi = {10.1146/annurev.astro.34.1.155},
       adsurl = {https://ui.adsabs.harvard.edu/abs/1996ARA&A..34..155B}
}

@ARTICLE{Beck2007,
       author = {{Beck}, R.},
        title = "{Magnetism in the spiral galaxy NGC 6946: magnetic arms, depolarization rings, dynamo modes, and helical fields}",
      journal = {\aap},
         year = 2007,
        month = aug,
       volume = {470},
       number = {2},
        pages = {539-556},
          doi = {10.1051/0004-6361:20066988},
archivePrefix = {arXiv},
       eprint = {0705.4163},
 primaryClass = {astro-ph},
       adsurl = {https://ui.adsabs.harvard.edu/abs/2007A&A...470..539B}
}

@ARTICLE{Law2011ApJ,
       author = {{Law}, C.~J. and {Gaensler}, B.~M. and {Bower}, G.~C. and {Backer}, D.~C. and {Bauermeister}, A. and {Croft}, S. and {Forster}, R. and {Gutierrez-Kraybill}, C. and {Harvey-Smith}, L. and {Heiles}, C. and {Hull}, C. and {Keating}, G. and {MacMahon}, D. and {Whysong}, D. and {Williams}, P.~K.~G. and {Wright}, M.},
        title = "{Spectropolarimetry with the Allen Telescope Array: Faraday Rotation Toward Bright Polarized Radio Galaxies}",
      journal = {\apj},
         year = 2011,
        month = feb,
       volume = {728},
       number = {1},
          eid = {57},
        pages = {57},
          doi = {10.1088/0004-637X/728/1/57},
archivePrefix = {arXiv},
       eprint = {1012.0945},
 primaryClass = {astro-ph.CO},
       adsurl = {https://ui.adsabs.harvard.edu/abs/2011ApJ...728...57L}
}

@INPROCEEDINGS{Heald2015,
       author = {{Heald}, George},
        title = "{Synchrotron Radiation and Faraday Rotation}",
    booktitle = {Magnetic Fields in Diffuse Media},
         year = 2015,
       editor = {{Lazarian}, Alexander and {de Gouveia Dal Pino}, Elisabete M. and {Melioli}, Claudio},
       series = {Astrophysics and Space Science Library},
       volume = {407},
        month = jan,
        pages = {41},
          doi = {10.1007/978-3-662-44625-6_3},
       adsurl = {https://ui.adsabs.harvard.edu/abs/2015ASSL..407...41H}
}

@ARTICLE{Durrer2013,
       author = {{Durrer}, Ruth and {Neronov}, Andrii},
        title = "{Cosmological magnetic fields: their generation, evolution and observation}",
      journal = {\aapr},
         year = 2013,
        month = jun,
       volume = {21},
          eid = {62},
        pages = {62},
          doi = {10.1007/s00159-013-0062-7},
archivePrefix = {arXiv},
       eprint = {1303.7121},
 primaryClass = {astro-ph.CO},
       adsurl = {https://ui.adsabs.harvard.edu/abs/2013A&ARv..21...62D}
}

@ARTICLE{Zweibel1997,
       author = {{Zweibel}, Ellen G. and {Lovelace}, Richard V.~E.},
        title = "{Stability of Magnetically Supported Disks}",
      journal = {\apj},
         year = 1997,
        month = jan,
       volume = {475},
       number = {1},
        pages = {260-262},
          doi = {10.1086/303530},
       adsurl = {https://ui.adsabs.harvard.edu/abs/1997ApJ...475..260Z}
}

@INPROCEEDINGS{Haverkorn2015,
       author = {{Haverkorn}, Marijke},
        title = "{Magnetic Fields in the Milky Way}",
    booktitle = {Magnetic Fields in Diffuse Media},
         year = 2015,
       editor = {{Lazarian}, Alexander and {de Gouveia Dal Pino}, Elisabete M. and {Melioli}, Claudio},
       series = {Astrophysics and Space Science Library},
       volume = {407},
        month = jan,
        pages = {483},
          doi = {10.1007/978-3-662-44625-6_17},
archivePrefix = {arXiv},
       eprint = {1406.0283},
 primaryClass = {astro-ph.GA},
       adsurl = {https://ui.adsabs.harvard.edu/abs/2015ASSL..407..483H}
}

@ARTICLE{Kovacs2024,
       author = {{Kovacs}, Timea Orsolya and {Mao}, Sui Ann and {Basu}, Aritra and {Ma}, Yik Ki and {Pakmor}, Ruediger and {Spitler}, Laura G. and {Walker}, Charles R.~H.},
        title = "{Dispersion and rotation measures from fast radio burst (FRB) host galaxies based on the TNG50 simulation}",
      journal = {\aap},
         year = 2024,
        month = oct,
       volume = {690},
          eid = {A47},
        pages = {A47},
          doi = {10.1051/0004-6361/202347459},
archivePrefix = {arXiv},
       eprint = {2407.16748},
 primaryClass = {astro-ph.GA},
       adsurl = {https://ui.adsabs.harvard.edu/abs/2024A&A...690A..47K}
}

@ARTICLE{Haverkorn2006,
       author = {{Haverkorn}, M. and {Gaensler}, B.~M. and {Brown}, J.~C. and {Bizunok}, N.~S. and {McClure-Griffiths}, N.~M. and {Dickey}, J.~M. and {Green}, A.~J.},
        title = "{Enhanced Small-Scale Faraday Rotation in the Galactic Spiral Arms}",
      journal = {\apjl},
         year = 2006,
        month = jan,
       volume = {637},
       number = {1},
        pages = {L33-L35},
          doi = {10.1086/500543},
archivePrefix = {arXiv},
       eprint = {astro-ph/0512456},
 primaryClass = {astro-ph},
       adsurl = {https://ui.adsabs.harvard.edu/abs/2006ApJ...637L..33H}
}

@ARTICLE{Haverkorn2008,
       author = {{Haverkorn}, M. and {Brown}, J.~C. and {Gaensler}, B.~M. and {McClure-Griffiths}, N.~M.},
        title = "{The Outer Scale of Turbulence in the Magnetoionized Galactic Interstellar Medium}",
      journal = {\apj},
         year = 2008,
        month = jun,
       volume = {680},
       number = {1},
        pages = {362-370},
          doi = {10.1086/587165},
archivePrefix = {arXiv},
       eprint = {0802.2740},
 primaryClass = {astro-ph},
       adsurl = {https://ui.adsabs.harvard.edu/abs/2008ApJ...680..362H}
}

@BOOK{Beck2013,
       author = {{Beck}, Rainer and {Balogh}, Andr{\'e} and {Bykov}, Andrei and {Treumann}, Rudolf A. and {Widrow}, Lawrence},
        title = "{Large-Scale Magnetic Fields in the Universe}",
         year = 2013,
          doi = {10.1007/978-1-4614-5728-2},
       adsurl = {https://ui.adsabs.harvard.edu/abs/2013lsmf.book.....B}
}

@ARTICLE{Zhang2022,
       author = {{Zhang}, M. and {Yuan}, Q. and {Liu}, J. -Y. and {Zhang}, L.},
        title = "{Role of the companion lensing galaxy in the CLASS gravitational lens B1152+199}",
      journal = {\mnras},
         year = 2022,
        month = mar,
       volume = {511},
       number = {1},
        pages = {1085-1103},
          doi = {10.1093/mnras/stac087},
archivePrefix = {arXiv},
       eprint = {2202.06926},
 primaryClass = {astro-ph.GA},
       adsurl = {https://ui.adsabs.harvard.edu/abs/2022MNRAS.511.1085Z}
}

@ARTICLE{Dyer1992,
       author = {{Dyer}, C.~C. and {Shaver}, Eric G.},
        title = "{On the Rotation of Polarization by a Gravitational Lens}",
      journal = {\apjl},
         year = 1992,
        month = may,
       volume = {390},
        pages = {L5},
          doi = {10.1086/186358},
       adsurl = {https://ui.adsabs.harvard.edu/abs/1992ApJ...390L...5D}
}

@ARTICLE{Tahani2018,
       author = {{Tahani}, M. and {Plume}, R. and {Brown}, J.~C. and {Kainulainen}, J.},
        title = "{Helical magnetic fields in molecular clouds?. A new method to determine the line-of-sight magnetic field structure in molecular clouds}",
      journal = {\aap},
         year = 2018,
        month = jun,
       volume = {614},
          eid = {A100},
        pages = {A100},
          doi = {10.1051/0004-6361/201732219},
archivePrefix = {arXiv},
       eprint = {1802.07831},
 primaryClass = {astro-ph.GA},
       adsurl = {https://ui.adsabs.harvard.edu/abs/2018A&A...614A.100T}
}

@ARTICLE{Reynolds1977,
       author = {{Reynolds}, R.~J.},
        title = "{Pulsar dispersion measures and Halpha emission measures: limits on the electron density and filling factor for the ionized interstellar gas.}",
      journal = {\apj},
         year = 1977,
        month = sep,
       volume = {216},
        pages = {433-439},
          doi = {10.1086/155484},
       adsurl = {https://ui.adsabs.harvard.edu/abs/1977ApJ...216..433R}
}

@ARTICLE{Manchester1981,
       author = {{Manchester}, R.~N. and {Taylor}, J.~H.},
        title = "{Observed and derived parameters for 330 pulsars.}",
      journal = {\aj},
         year = 1981,
        month = dec,
       volume = {86},
        pages = {1953-1973},
          doi = {10.1086/113078},
       adsurl = {https://ui.adsabs.harvard.edu/abs/1981AJ.....86.1953M}
}

@ARTICLE{Taylor1993,
       author = {{Taylor}, J.~H. and {Cordes}, J.~M.},
        title = "{Pulsar Distances and the Galactic Distribution of Free Electrons}",
      journal = {\apj},
         year = 1993,
        month = jul,
       volume = {411},
        pages = {674},
          doi = {10.1086/172870},
       adsurl = {https://ui.adsabs.harvard.edu/abs/1993ApJ...411..674T}
}

@ARTICLE{Brentjens2005,
       author = {{Brentjens}, M.~A. and {de Bruyn}, A.~G.},
        title = "{Faraday rotation measure synthesis}",
      journal = {\aap},
         year = 2005,
        month = oct,
       volume = {441},
       number = {3},
        pages = {1217-1228},
          doi = {10.1051/0004-6361:20052990},
archivePrefix = {arXiv},
       eprint = {astro-ph/0507349},
 primaryClass = {astro-ph},
       adsurl = {https://ui.adsabs.harvard.edu/abs/2005A&A...441.1217B}
}

@ARTICLE{Burn1966,
       author = {{Burn}, B.~J.},
        title = "{On the depolarization of discrete radio sources by Faraday dispersion}",
      journal = {\mnras},
         year = 1966,
        month = jan,
       volume = {133},
        pages = {67},
          doi = {10.1093/mnras/133.1.67},
       adsurl = {https://ui.adsabs.harvard.edu/abs/1966MNRAS.133...67B}
}

@ARTICLE{MacLow2004,
       author = {{Mac Low}, Mordecai-Mark and {Klessen}, Ralf S.},
        title = "{Control of star formation by supersonic turbulence}",
      journal = {Reviews of Modern Physics},
         year = 2004,
        month = jan,
       volume = {76},
       number = {1},
        pages = {125-194},
          doi = {10.1103/RevModPhys.76.125},
archivePrefix = {arXiv},
       eprint = {astro-ph/0301093},
 primaryClass = {astro-ph},
       adsurl = {https://ui.adsabs.harvard.edu/abs/2004RvMP...76..125M}
}

@ARTICLE{Klessen2016,
       author = {{Klessen}, Ralf S. and {Glover}, Simon C.~O.},
        title = "{Physical Processes in the Interstellar Medium}",
      journal = {Saas-Fee Advanced Course},
         year = 2016,
        month = jan,
       volume = {43},
        pages = {85},
          doi = {10.1007/978-3-662-47890-5_2},
archivePrefix = {arXiv},
       eprint = {1412.5182},
 primaryClass = {astro-ph.GA},
       adsurl = {https://ui.adsabs.harvard.edu/abs/2016SAAS...43...85K}
}

@ARTICLE{McKee2007,
       author = {{McKee}, Christopher F. and {Ostriker}, Eve C.},
        title = "{Theory of Star Formation}",
      journal = {\araa},
         year = 2007,
        month = sep,
       volume = {45},
       number = {1},
        pages = {565-687},
          doi = {10.1146/annurev.astro.45.051806.110602},
archivePrefix = {arXiv},
       eprint = {0707.3514},
 primaryClass = {astro-ph},
       adsurl = {https://ui.adsabs.harvard.edu/abs/2007ARA&A..45..565M}
}

@ARTICLE{Schober2012a,
       author = {{Schober}, Jennifer and {Schleicher}, Dominik and {Federrath}, Christoph and {Klessen}, Ralf and {Banerjee}, Robi},
        title = "{Magnetic field amplification by small-scale dynamo action: Dependence on turbulence models and Reynolds and Prandtl numbers}",
      journal = {\pre},
         year = 2012,
        month = feb,
       volume = {85},
       number = {2},
          eid = {026303},
        pages = {026303},
          doi = {10.1103/PhysRevE.85.026303},
archivePrefix = {arXiv},
       eprint = {1109.4571},
 primaryClass = {astro-ph.CO},
       adsurl = {https://ui.adsabs.harvard.edu/abs/2012PhRvE..85b6303S}
}

@ARTICLE{Schober2012b,
       author = {{Schober}, Jennifer and {Schleicher}, Dominik and {Federrath}, Christoph and {Glover}, Simon and {Klessen}, Ralf S. and {Banerjee}, Robi},
        title = "{The Small-scale Dynamo and Non-ideal Magnetohydrodynamics in Primordial Star Formation}",
      journal = {\apj},
         year = 2012,
        month = aug,
       volume = {754},
       number = {2},
          eid = {99},
        pages = {99},
          doi = {10.1088/0004-637X/754/2/99},
archivePrefix = {arXiv},
       eprint = {1204.0658},
 primaryClass = {astro-ph.CO},
       adsurl = {https://ui.adsabs.harvard.edu/abs/2012ApJ...754...99S}
}

@ARTICLE{Schober2013,
       author = {{Schober}, J. and {Schleicher}, D.~R.~G. and {Klessen}, R.~S.},
        title = "{Magnetic field amplification in young galaxies}",
      journal = {\aap},
         year = 2013,
        month = dec,
       volume = {560},
          eid = {A87},
        pages = {A87},
          doi = {10.1051/0004-6361/201322185},
archivePrefix = {arXiv},
       eprint = {1310.0853},
 primaryClass = {astro-ph.GA},
       adsurl = {https://ui.adsabs.harvard.edu/abs/2013A&A...560A..87S}
}

@ARTICLE{Schober2015,
       author = {{Schober}, J. and {Schleicher}, D.~R.~G. and {Federrath}, C. and {Bovino}, S. and {Klessen}, R.~S.},
        title = "{Saturation of the turbulent dynamo}",
      journal = {\pre},
         year = 2015,
        month = aug,
       volume = {92},
       number = {2},
          eid = {023010},
        pages = {023010},
          doi = {10.1103/PhysRevE.92.023010},
archivePrefix = {arXiv},
       eprint = {1506.02182},
 primaryClass = {physics.plasm-ph},
       adsurl = {https://ui.adsabs.harvard.edu/abs/2015PhRvE..92b3010S}
}

@ARTICLE{Federrath2011,
       author = {{Federrath}, C. and {Chabrier}, G. and {Schober}, J. and {Banerjee}, R. and {Klessen}, R.~S. and {Schleicher}, D.~R.~G.},
        title = "{Mach Number Dependence of Turbulent Magnetic Field Amplification: Solenoidal versus Compressive Flows}",
      journal = {\prl},
         year = 2011,
        month = sep,
       volume = {107},
       number = {11},
          eid = {114504},
        pages = {114504},
          doi = {10.1103/PhysRevLett.107.114504},
archivePrefix = {arXiv},
       eprint = {1109.1760},
 primaryClass = {physics.flu-dyn},
       adsurl = {https://ui.adsabs.harvard.edu/abs/2011PhRvL.107k4504F}
}

@ARTICLE{Federrath2012,
       author = {{Federrath}, Christoph and {Klessen}, Ralf S.},
        title = "{The Star Formation Rate of Turbulent Magnetized Clouds: Comparing Theory, Simulations, and Observations}",
      journal = {\apj},
         year = 2012,
        month = dec,
       volume = {761},
       number = {2},
          eid = {156},
        pages = {156},
          doi = {10.1088/0004-637X/761/2/156},
archivePrefix = {arXiv},
       eprint = {1209.2856},
 primaryClass = {astro-ph.SR},
       adsurl = {https://ui.adsabs.harvard.edu/abs/2012ApJ...761..156F}
}

@ARTICLE{Federrath2015,
       author = {{Federrath}, Christoph},
        title = "{Inefficient star formation through turbulence, magnetic fields and feedback}",
      journal = {\mnras},
         year = 2015,
        month = jul,
       volume = {450},
       number = {4},
        pages = {4035-4042},
          doi = {10.1093/mnras/stv941},
archivePrefix = {arXiv},
       eprint = {1504.03690},
 primaryClass = {astro-ph.SR},
       adsurl = {https://ui.adsabs.harvard.edu/abs/2015MNRAS.450.4035F}
}

@ARTICLE{Schleicher2010,
       author = {{Schleicher}, D.~R.~G. and {Banerjee}, R. and {Sur}, S. and {Arshakian}, T.~G. and {Klessen}, R.~S. and {Beck}, R. and {Spaans}, M.},
        title = "{Small-scale dynamo action during the formation of the first stars and galaxies. I. The ideal MHD limit}",
      journal = {\aap},
         year = 2010,
        month = nov,
       volume = {522},
          eid = {A115},
        pages = {A115},
          doi = {10.1051/0004-6361/201015184},
archivePrefix = {arXiv},
       eprint = {1003.1135},
 primaryClass = {astro-ph.CO},
       adsurl = {https://ui.adsabs.harvard.edu/abs/2010A&A...522A.115S}
}

@ARTICLE{Brandenburg2005,
       author = {{Brandenburg}, Axel and {Subramanian}, Kandaswamy},
        title = "{Astrophysical magnetic fields and nonlinear dynamo theory}",
      journal = {\physrep},
         year = 2005,
        month = oct,
       volume = {417},
       number = {1-4},
        pages = {1-209},
          doi = {10.1016/j.physrep.2005.06.005},
archivePrefix = {arXiv},
       eprint = {astro-ph/0405052},
 primaryClass = {astro-ph},
       adsurl = {https://ui.adsabs.harvard.edu/abs/2005PhR...417....1B}
}

@ARTICLE{Schekochihin2004,
       author = {{Schekochihin}, Alexander A. and {Cowley}, Steven C. and {Taylor}, Samuel F. and {Maron}, Jason L. and {McWilliams}, James C.},
        title = "{Simulations of the Small-Scale Turbulent Dynamo}",
      journal = {\apj},
         year = 2004,
        month = sep,
       volume = {612},
       number = {1},
        pages = {276-307},
          doi = {10.1086/422547},
archivePrefix = {arXiv},
       eprint = {astro-ph/0312046},
 primaryClass = {astro-ph},
       adsurl = {https://ui.adsabs.harvard.edu/abs/2004ApJ...612..276S}
}

@ARTICLE{Sur2010,
       author = {{Sur}, Sharanya and {Schleicher}, D.~R.~G. and {Banerjee}, Robi and {Federrath}, Christoph and {Klessen}, Ralf S.},
        title = "{The Generation of Strong Magnetic Fields During the Formation of the First Stars}",
      journal = {\apjl},
         year = 2010,
        month = oct,
       volume = {721},
       number = {2},
        pages = {L134-L138},
          doi = {10.1088/2041-8205/721/2/L134},
archivePrefix = {arXiv},
       eprint = {1008.3481},
 primaryClass = {astro-ph.CO},
       adsurl = {https://ui.adsabs.harvard.edu/abs/2010ApJ...721L.134S}
}

@ARTICLE{Federrath2014,
       author = {{Federrath}, Christoph and {Schober}, Jennifer and {Bovino}, Stefano and {Schleicher}, Dominik R.~G.},
        title = "{The Turbulent Dynamo in Highly Compressible Supersonic Plasmas}",
      journal = {\apjl},
         year = 2014,
        month = dec,
       volume = {797},
       number = {2},
          eid = {L19},
        pages = {L19},
          doi = {10.1088/2041-8205/797/2/L19},
archivePrefix = {arXiv},
       eprint = {1411.4707},
 primaryClass = {astro-ph.GA},
       adsurl = {https://ui.adsabs.harvard.edu/abs/2014ApJ...797L..19F}
}

@ARTICLE{Rappaz2024,
       author = {{Rappaz}, Y. and {Schober}, J. and {Bendre}, A.~B. and {Seta}, A. and {Federrath}, C.},
        title = "{Cosmic evolution of the Faraday rotation measure in the intracluster medium of galaxy clusters}",
      journal = {\aap},
         year = 2024,
        month = nov,
       volume = {691},
          eid = {A132},
        pages = {A132},
          doi = {10.1051/0004-6361/202451119},
archivePrefix = {arXiv},
       eprint = {2409.18580},
 primaryClass = {astro-ph.CO},
       adsurl = {https://ui.adsabs.harvard.edu/abs/2024A&A...691A.132R}
}



\appendix

\section{All maps}
\label{app:maps}

\begin{figure*}[h!]
	\centering
	\includegraphics[width=0.95\textwidth]{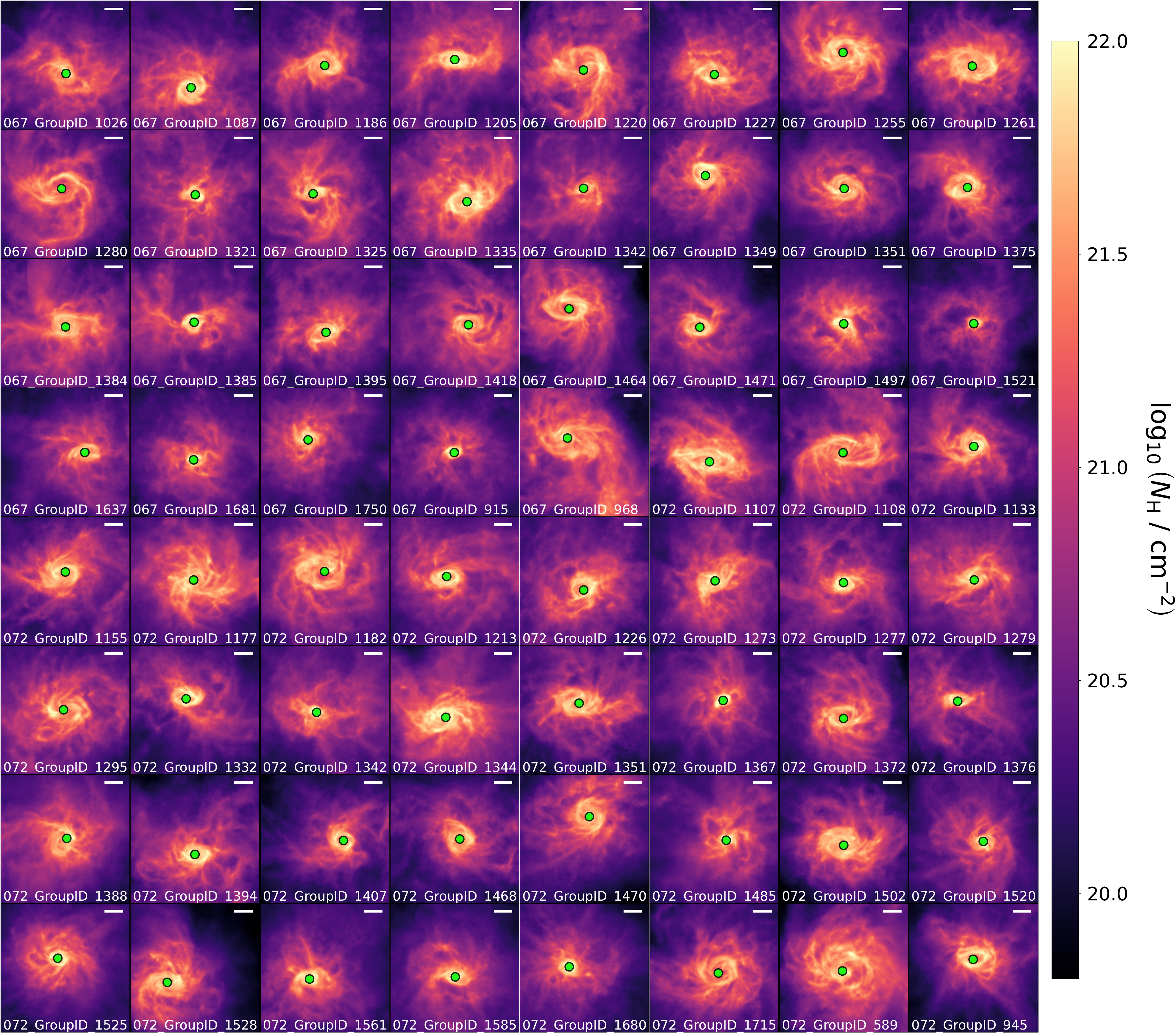}
	\caption{The gas column density maps $N_{\mathrm{H}}$ for all considered TNG50 galaxies for an inclination of $i=45^\circ$. Red dots indicate the designated center of gas mass of each galaxy. Galaxies with an ID of 067 are at a redshift of $z=0.4$ and an ID of 072 a redshift of $z=0.5$. White lines in the upper right corner of each panel represent a length of $10\ \mathrm{kpc}$.  }
	\label{fig:NH_maps}
\end{figure*}
\begin{figure*}[h!]
	\centering
	\includegraphics[width=0.95\textwidth]{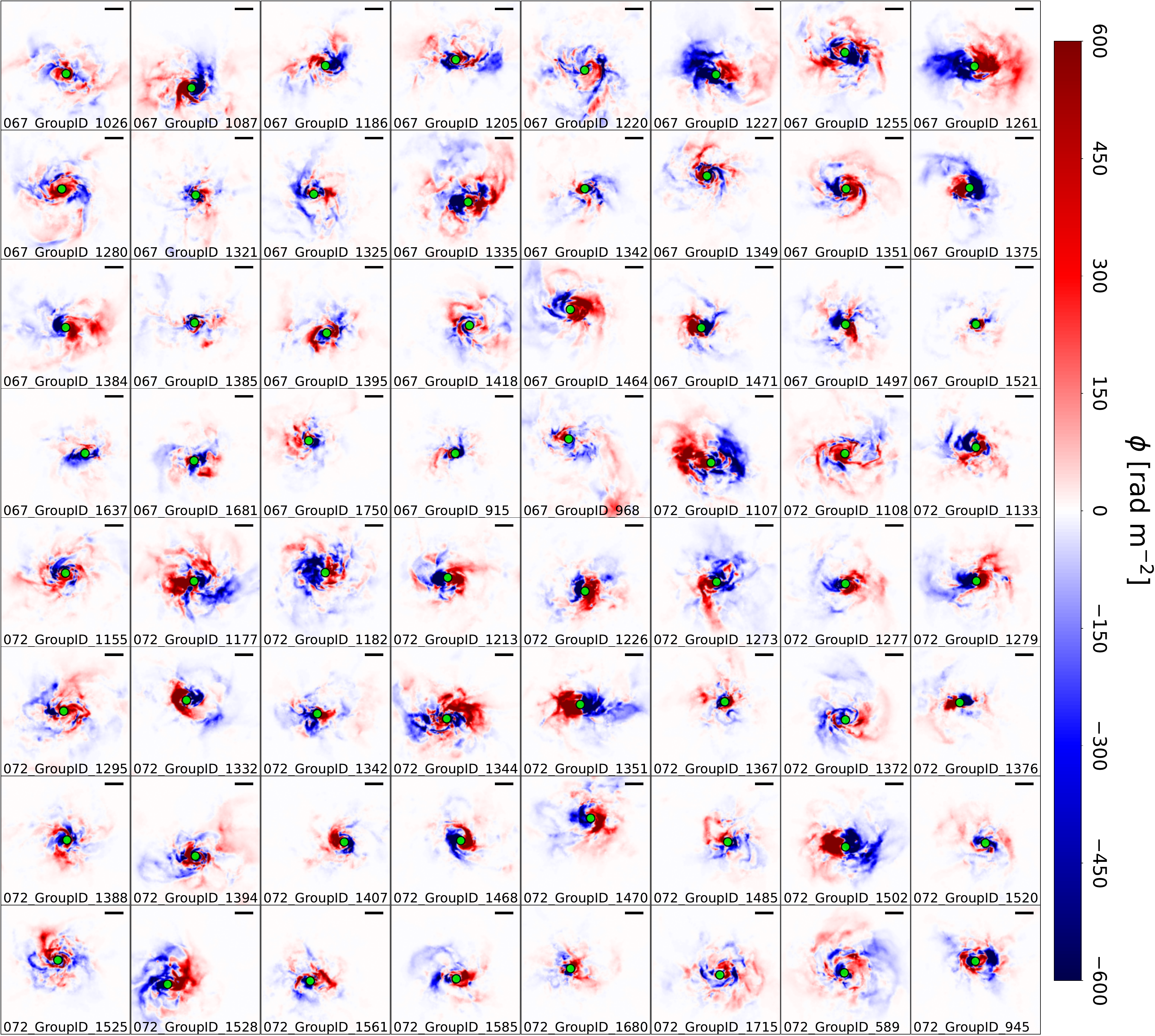}
	\caption{The same as Fig.~\ref{fig:NH_maps} but for the integrated Faraday  depth $\phi$.}
	\label{fig:RM_maps}
\end{figure*}
In this section, we provide maps of the gas column density $N_{\mathrm{H}}$ in Fig.\ref{fig:NH_maps} and the Faraday depth $\phi$, directly integrated through the simulation domain, in Fig.\ref{fig:RM_maps}.

\end{document}